%% file: main.tex
\documentclass[review, 3p, twocolumn]{elsarticle}

\usepackage{graphicx}% Include figure files
\usepackage{wrapfig}
\usepackage{bm}% bold math
\usepackage{hyperref}% add hypertext capabilities
\usepackage{siunitx}
\usepackage{miller}
\usepackage[version=4]{mhchem} % for chemical formulas
\usepackage{multirow}
\usepackage{tabularx}
\usepackage[singlespacing]{setspace}

\ExplSyntaxOn

\NewDocumentCommand{\hmn}{m}
 {
  \ensuremath
   {
    \hermannmauguin_group:n { #1 }
   }
 }

\tl_new:N \l_hermannmauguin_input_tl
\tl_new:N \l_hermannmauguin_output_tl

\cs_new_protected:Nn \hermannmauguin_group:n
 {
  \tl_set:Nn \l_hermannmauguin_input_tl { #1 }
  \tl_clear:N \l_hermannmauguin_output_tl
  \tl_map_inline:Nn \l_hermannmauguin_input_tl
   {
    \__hermannmauguin_item:n { ##1 }
   }
  \! % kill the first \,
  \tl_use:N \l_hermannmauguin_output_tl
 }

\cs_new_protected:Nn \__hermannmauguin_item:n
 {
  \str_case:nnF { #1 }
   {
    {*}{ \__hermannmauguin_put:n { \__hermannmauguin_inverse:Nn } }
    {-}{ \__hermannmauguin_put:n { \__hermannmauguin_overline:Nn } }
    {i}{ \__hermannmauguin_put:n { \,\infty } }
   }
   { \__hermannmauguin_put:n { \, {#1} } }
 }

\cs_new_protected:Nn \__hermannmauguin_put:n
 {
  \tl_put_right:Nn \l_hermannmauguin_output_tl { #1 }
 }

\cs_new_protected:Nn \__hermannmauguin_overline:Nn
 {% #1 should be \,; #2 is the number to operate on
  #1 \mkern1mu\overline{\mkern-1mu#2\mkern-1mu}\mkern1mu
 }
\cs_new_protected:Nn \__hermannmauguin_inverse:Nn
 {% #1 should be \,; #2 is the number to operate on
  #1 \frac{ #2 } { m }
 }

\ExplSyntaxOff

\begin{document}

\begin{frontmatter}

%\title{Thickness Dependence of Orbital Mapping}
\title{Optimizing Experimental Parameters for Orbital Mapping}
%\title{Orbital Mapping: Glorious Future or Dead on Arrival?

\author[1]{Manuel Ederer}
\ead{manuel.ederer@tuwien.ac.at}

\author[1]{Stefan L\"{o}ffler}
\address[1]{University  Service  Centre  for  Transmission  Electron  Microscopy, TU Wien, Wiedner Hauptstraße 8-10/E057-02, 1040 Wien, Austria}

\date{\today}% It is always \today, today,

\begin{abstract}
\noindent
A new material characterization technique is emerging for the transmission electron microscope (TEM).
Using electron energy-loss spectroscopy, real space mappings of the underlying electronic transitions in the sample, so called orbital maps, can be produced.
Thus, unprecedented insight into the electronic orbitals responsible for most of the electrical, magnetic and optical properties of bulk materials can be gained.
However, the incredibly demanding requirements on spatial as well as spectral resolution paired with the low signal-to-noise ratio severely limits the day-to-day use of this new technique.
With the use of simulations, we strive to alleviate these challenges as much as possible by identifying optimal experimental parameters.
In this manner, we investigate representative examples of a transition metal oxide, a material consisting entirely of light elements, and an interface between two different materials to find and compare acceptable ranges for sample thickness, acceleration voltage and electron dose for a scanning probe as well as for parallel illumination.

\end{abstract}


\end{frontmatter}

\input{introduction}

\input{methods}

\input{rutile_STEM}

\input{rutile_TEM}
\input{graphite}
\input{STOLMO}

\section{Discussion and Conclusion}
In this work, we have successfully applied an image difference metric to energy-filtered (S)TEM images with the goal of identifying optimal sets of parameters for imaging.
The specific materials, i.e., rutile, graphite and a heterostructure of \ce{SrTiO3}-\ce{LaMnO3}, and the energy windows have been selected such that underlying electronic transitions can be mapped in real space.
Thus, in principle, these orbital maps allow to directly measure the spatial dependence of initially unoccupied orbitals above Fermi energy.
We have accomplished the task of finding suitable sets of parameters by calculating and minimizing the image difference between a perfect, albeit unachievable, image and the image for the specific set of parameters.\\
In the process of finding the optimal, or at least acceptable, sample thickness of the chosen materials, we have encountered a fundamental inequality between EFTEM images and STEM spectral images concerning the dependence of the image difference on sample thickness.
While the image difference of the STEM orbital maps increases fairly monotonically with the thickness, the EFTEM orbital maps display a periodicity which leads in many cases to comparably thicker samples being still acceptable.
Further, in almost all the investigated cases, a slight defocus of $-2$~nm to $-5$~nm further decreased the EFTEM image difference.
It appears that, in principle, EFTEM has the advantage over STEM concerning the difficulty and strictness in sample preparation.
However, this can only be certainly said for the assumed ideal imaging conditions, i.e. no aberrations or other lens errors, no sample drift or damage and no noise beside shot noise.
Under more realistic conditions the preference could easily shift towards STEM, as it is in general easier for STEM to apply post-processing in order to mitigate non-ideal imaging conditions.
Otherwise, we do not find a fundamental preference for either EFTEM or STEM in regards of orbital mapping.\\
In conclusion, orbital mapping, although not universally applicable (yet), proves to be a valuable tool when employed with meticulous planning and extensive theoretical simulations.
For each of the materials investigated, we find certain parameter ranges together with a realistically achievable sample thickness that result in images with distinctly recognizable orbital shapes.
Even when a finite electron dose and shot noise is considered, orbital mapping promises to provide valuable insights into the electronic structure and chemical binding of specific materials.

\noindent
We acknowledge financial support by the Austrian Science Fund (FWF) under grant number I4309-N36.
Further, we acknowledge TU Wien Bibliothek for financial support through its Open Access Funding Programme.
The computational results presented were achieved, in part, using the Vienna Scientific Cluster (VSC).\\

\bibliography{TEM, metric, mats}% Produces the bibliography via BibTeX.
\bibliographystyle{elsarticle-num}

\end{document}

%\preprint{APS/123-QED}

\begin{frontmatter}

\title{Supplementary Information for: "Optimizing Experimental Parameters for Orbital Mapping"}

\author[1]{Manuel Ederer}
\ead{manuel.ederer@tuwien.ac.at}

\author[1]{Stefan L\"{o}ffler}
\address[1]{University  Service  Centre  for  Transmission  Electron  Microscopy, TU Wien, Wiedner Hauptstraße 8-10/E057-02, 1040 Wien, Austria}

\date{\today}% It is always \today, today,

\end{frontmatter}

\externaldocument{main}
\renewcommand{\thefigure}{S1}
\begin{figure*}
    \centering
    \includegraphics{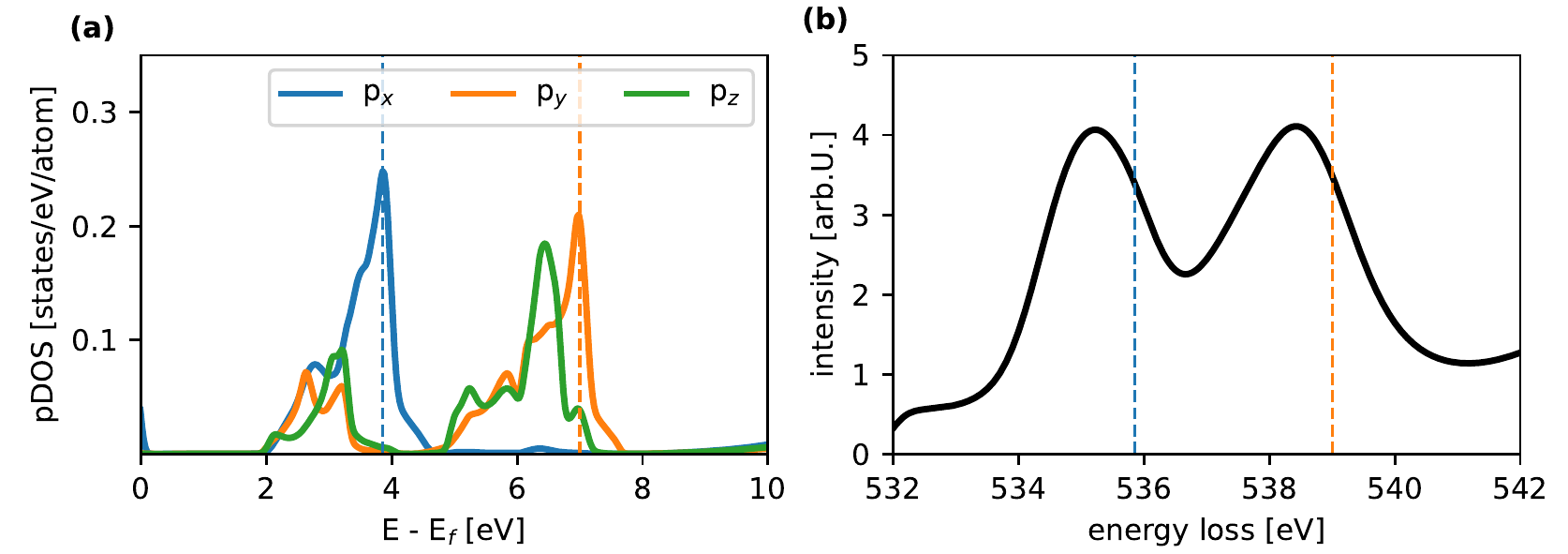}
    \caption{(a) DOS of rutile \ce{O}.
             (b) Energy loss near edge spectrum of the \ce{O} K-edge of rutile.
             Dashed lines mark the energy loss investigated in the main text.
}
\end{figure*}

\renewcommand{\thefigure}{S2}
\begin{figure*}
    \centering
    \includegraphics{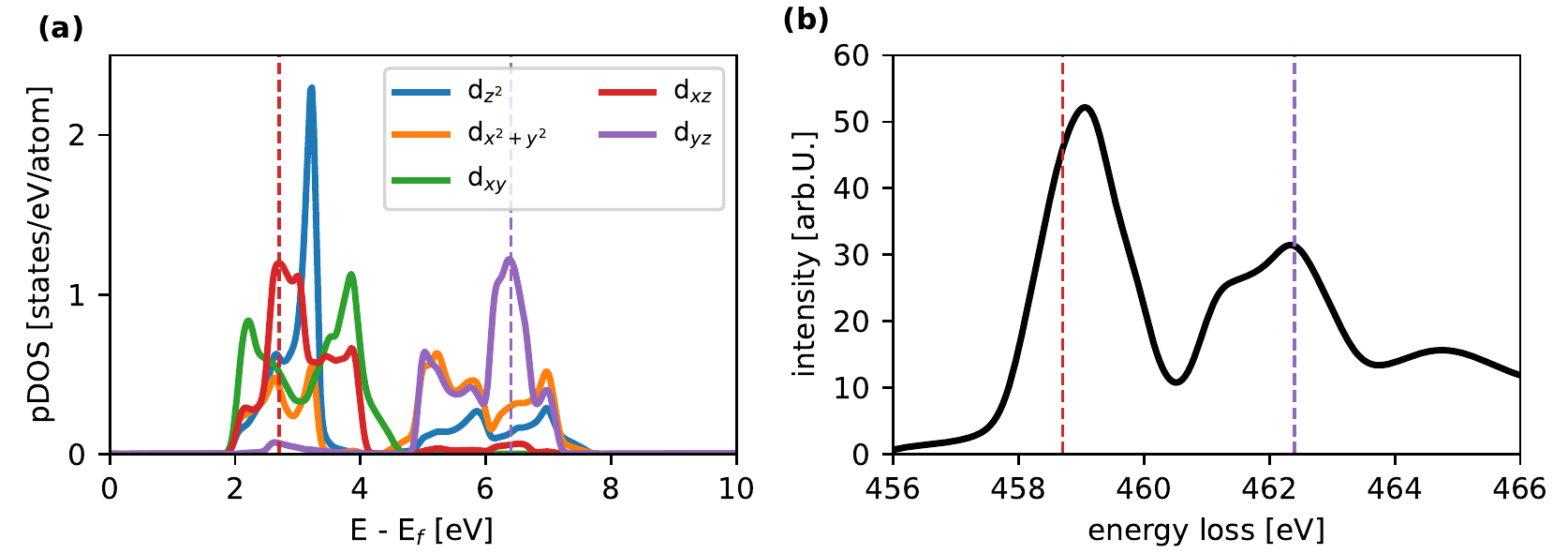}
    \caption{(a) DOS of rutile \ce{Ti}.
             (b) Energy loss near edge spectrum of the \ce{Ti} L$_3$-edge of rutile.
             Dashed lines mark the energy loss investigated in the main text.
}
\end{figure*}

\renewcommand{\thefigure}{S3}
\begin{figure*}
    \centering
    \includegraphics{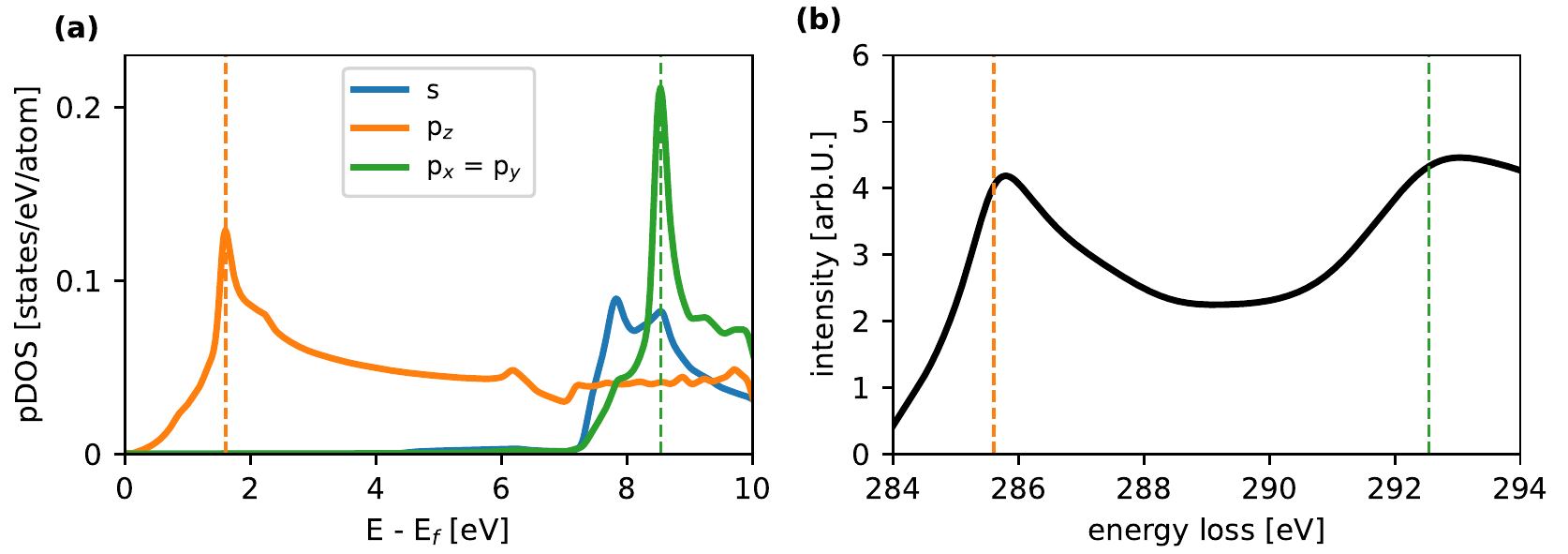}
    \caption{(a) DOS of graphite \ce{C}.
             (b) Energy loss near edge spectrum of the \ce{C} K-edge of graphite.
             Dashed lines mark the energy loss investigated in the main text.
}
\end{figure*}

\renewcommand{\thefigure}{S4}
\begin{figure*}
    \centering
    \includegraphics{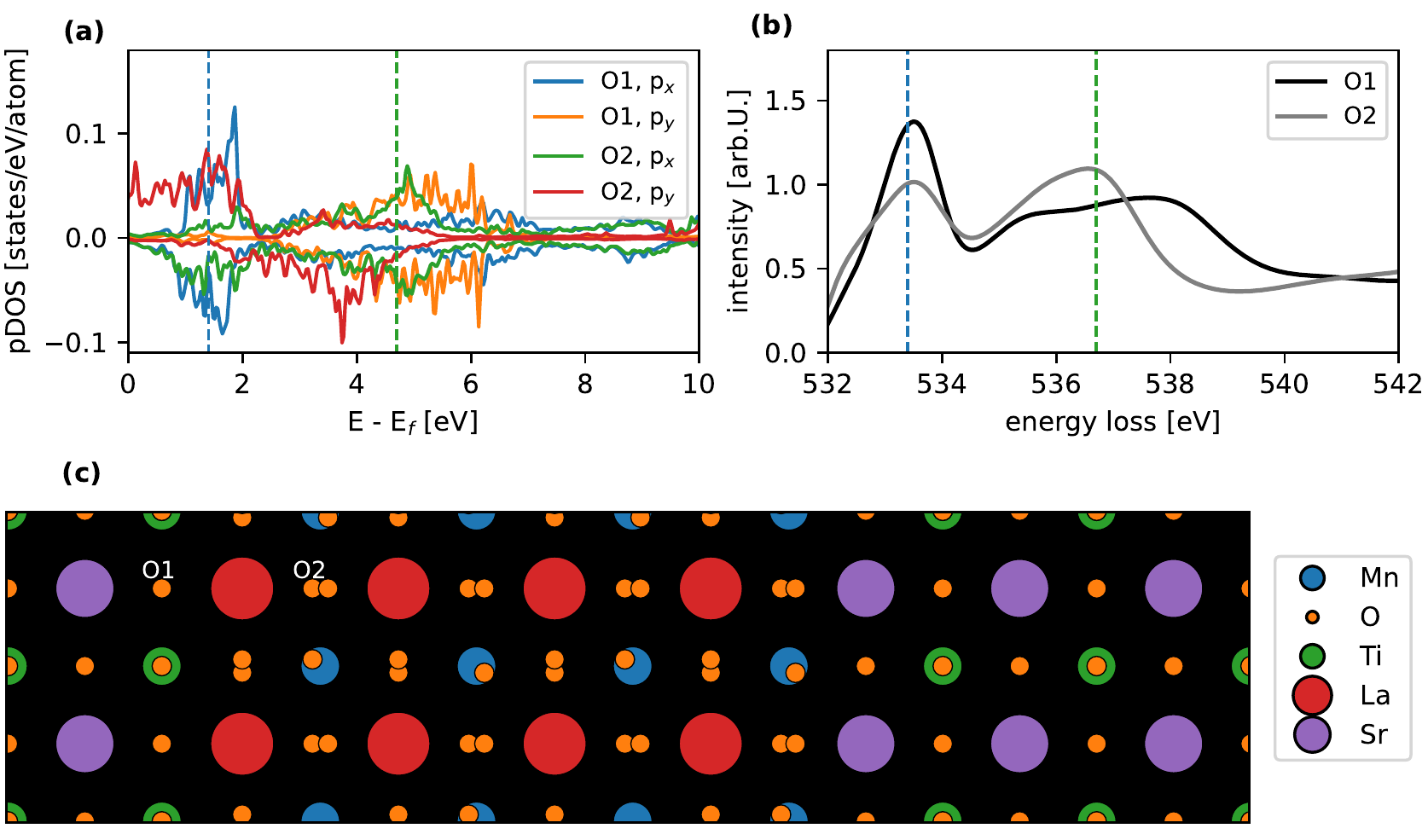}
    \caption{(a) Spin polarized DOS of two select \ce{O} atoms in a \ce{SrTiO3}-\ce{LaMnO3} interface.
             (b) Energy loss near edge spectrum of the \ce{O} K-edge of the two select \ce{O} atoms.
             Dashed lines mark the energy loss investigated in the main text.
             (c) Schematic of the \ce{SrTiO3}-\ce{LaMnO3} interface in the \hkl[010] projection.
             The positions of the \ce{O} atoms presented in (a) and (b) are marked with \ce{O}1 and \ce{O}2.
}
    \label{SI_MSE}
\end{figure*}

\renewcommand{\thefigure}{S5}
\begin{figure*}
    \centering
    \includegraphics{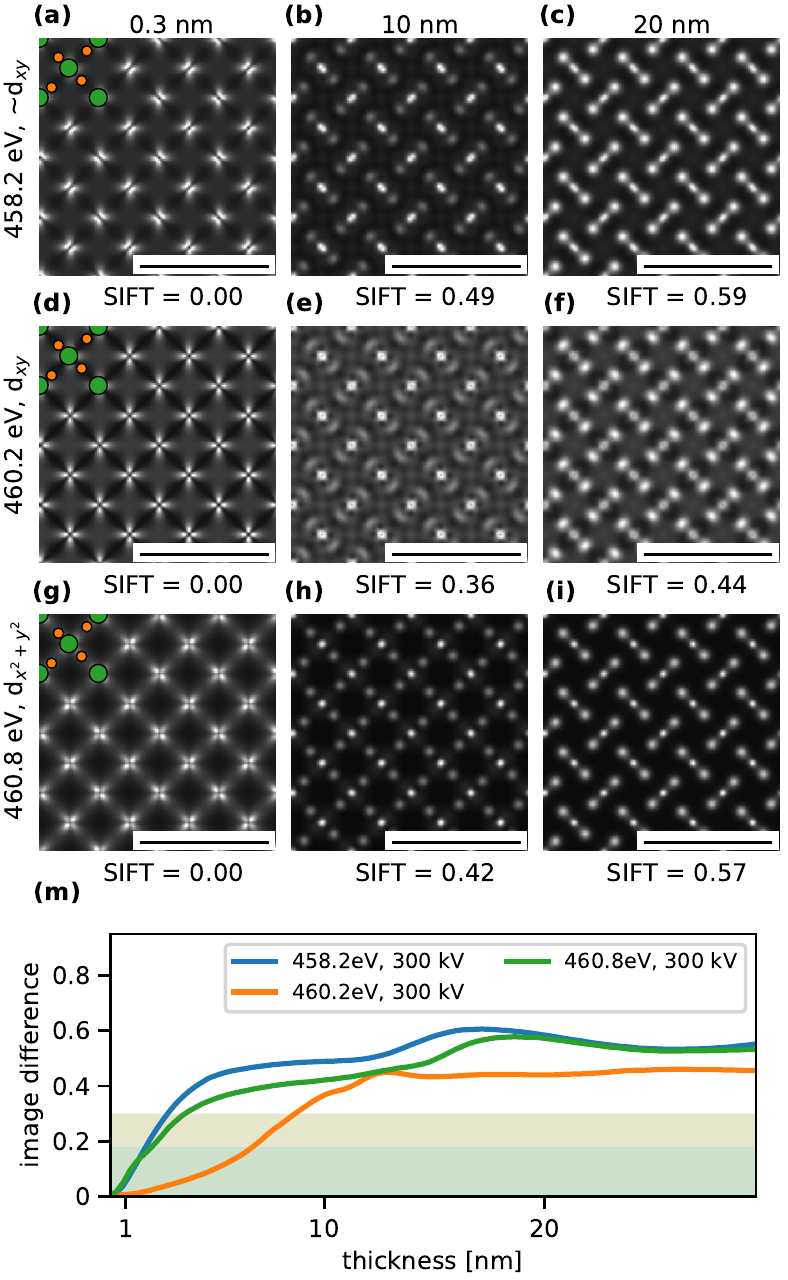}
    \caption{(a) - (l) Simulated energy-filtered TEM images of rutile in the \hkl[0 0 1] zone axis for the indicated sample thickness and energy loss.
    All scale bars indicate 1~nm. 
    Green and orange dots mark the positions of \ce{Ti} and \ce{O} atoms, respectively.
    (m) SIFT image difference as function of sample thickness.
    All image differences are relative to the respective energy-filtered image for 0.3~nm sample thickness and 300~kV acceleration voltage.  
    }
\end{figure*}

\renewcommand{\thefigure}{S6}
\begin{figure*}
    \centering
    \includegraphics{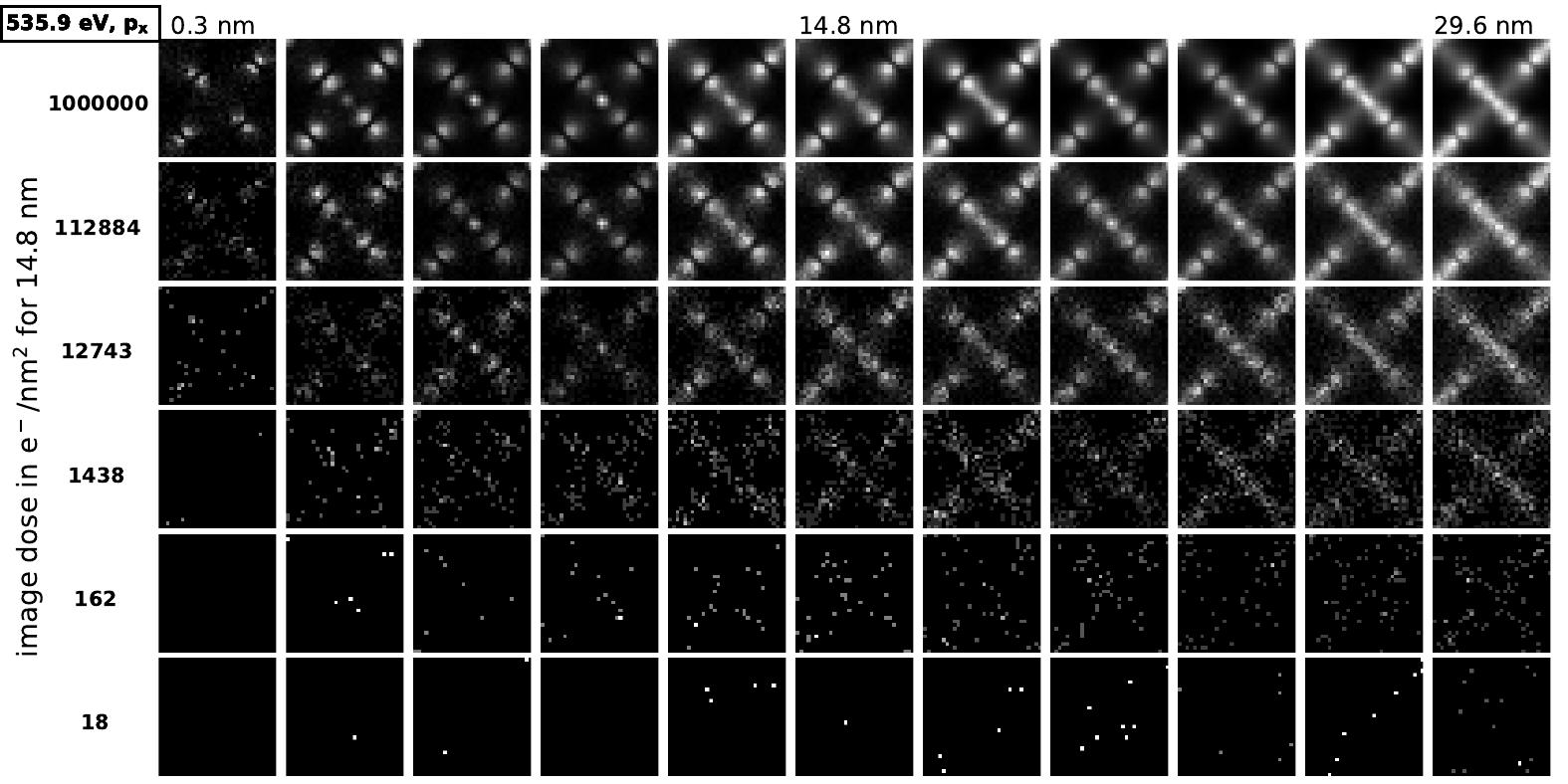}
    \caption{Simulated EFTEM images of a single rutile unit cell with added shot noise for varying sample thickness.
    The image dose values are given for the 14.8~nm thick sample.
    For the other cases, the image dose is scaled with the norm of the specific infinite dose image in order to simulate a constant incident dose.
}
\end{figure*}

\renewcommand{\thefigure}{S7}
\begin{figure*}
    \centering
    \includegraphics{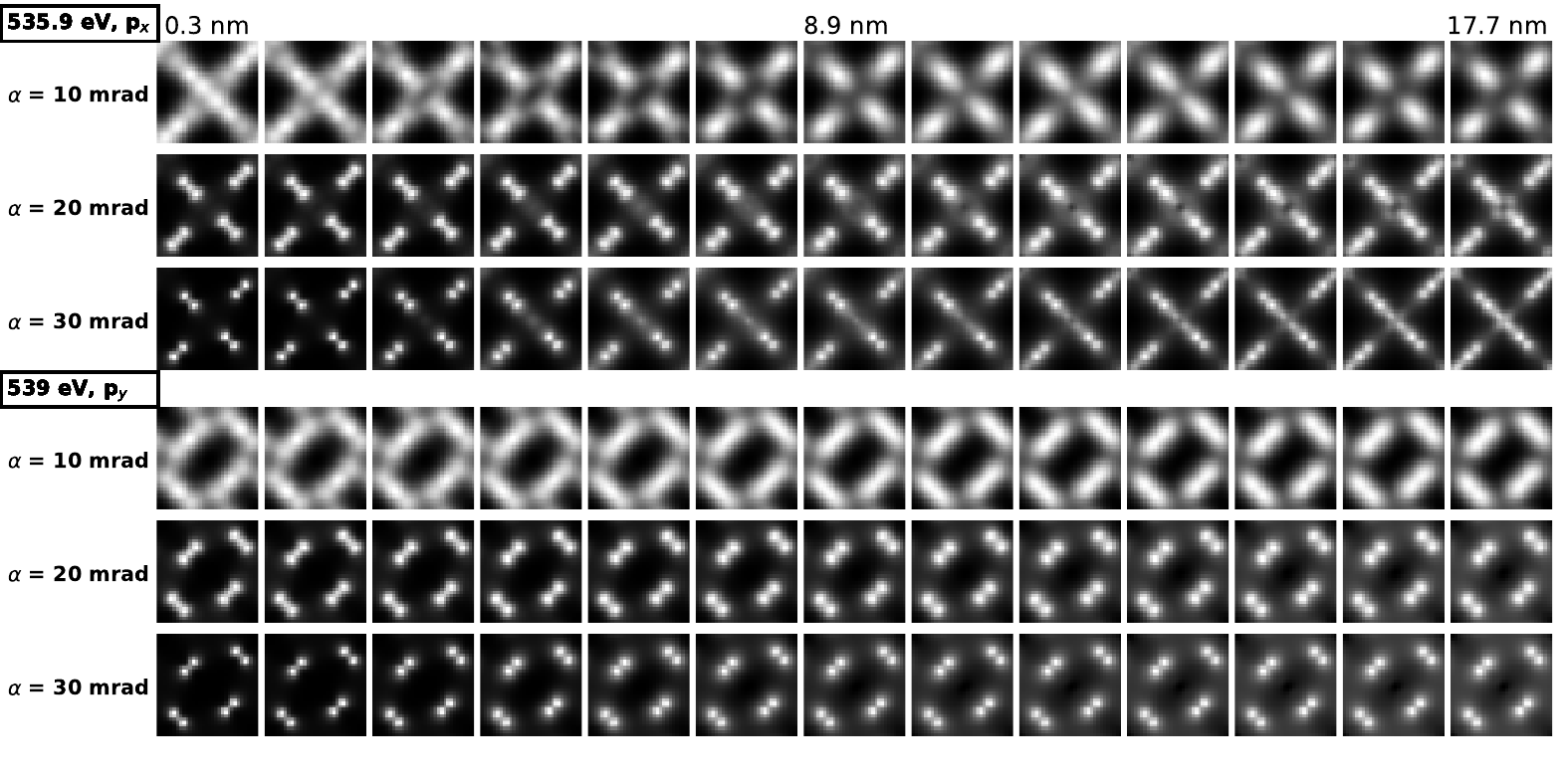}
    \caption{Simulated STEM-EELS maps of a single rutile unit cell for varying sample thicknesses and convergence semi-angles $\alpha$. 
   A collection semi-angle $\beta$ of 50~mrad was used for all images.
}
\end{figure*}

\renewcommand{\thefigure}{S8}
\begin{figure*}
    \centering
    \includegraphics{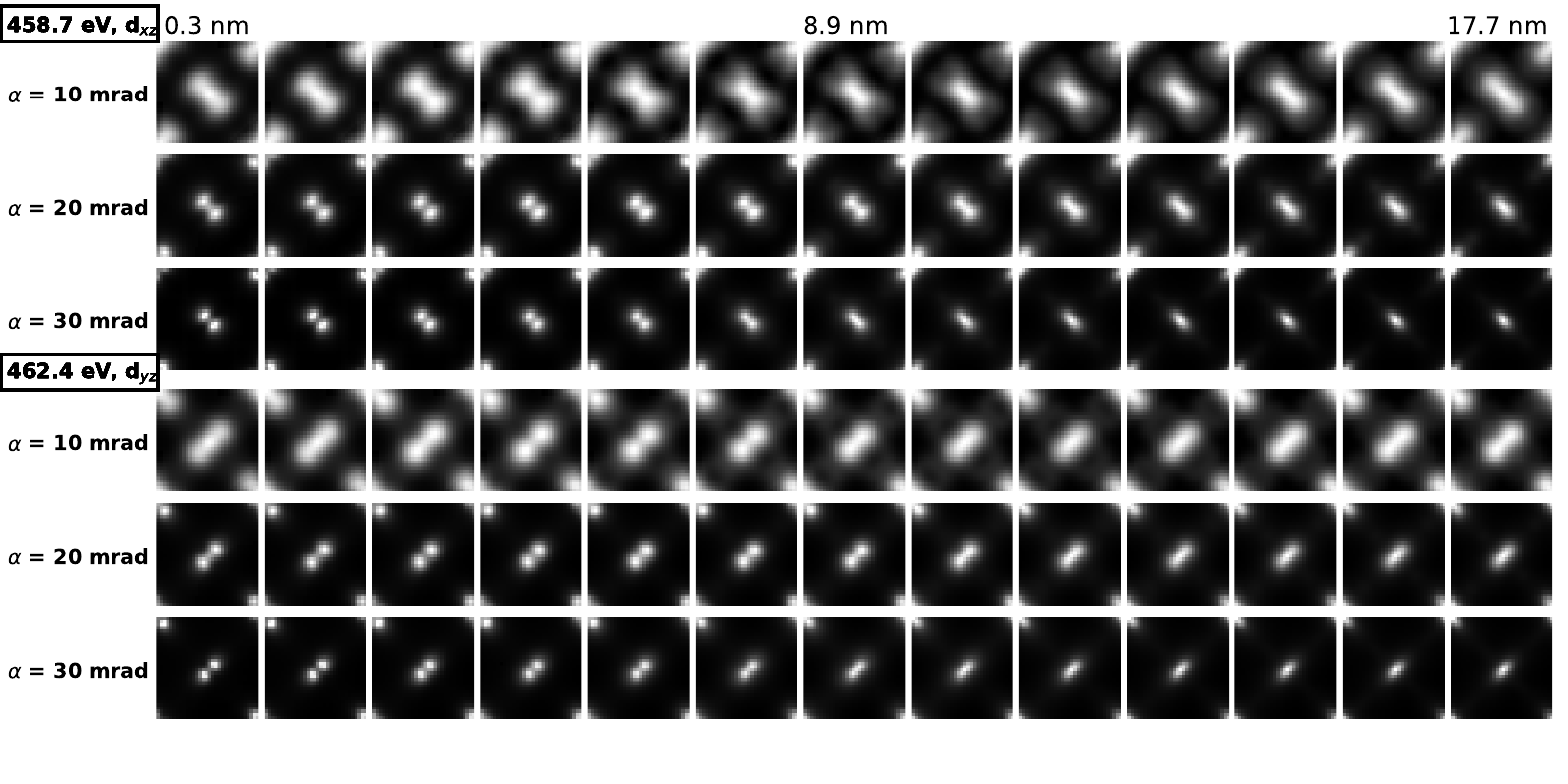}
    \caption{Simulated STEM-EELS maps of a single rutile unit cell for varying sample thicknesses and convergence semi-angles $\alpha$. 
   A collection semi-angle $\beta$ of 50~mrad was used for all images.
   }
    \label{fig_STEM_rutile_convanglewithpix_Ti}
\end{figure*}

\renewcommand{\thefigure}{S9}
\begin{figure*}
    \centering
    \includegraphics{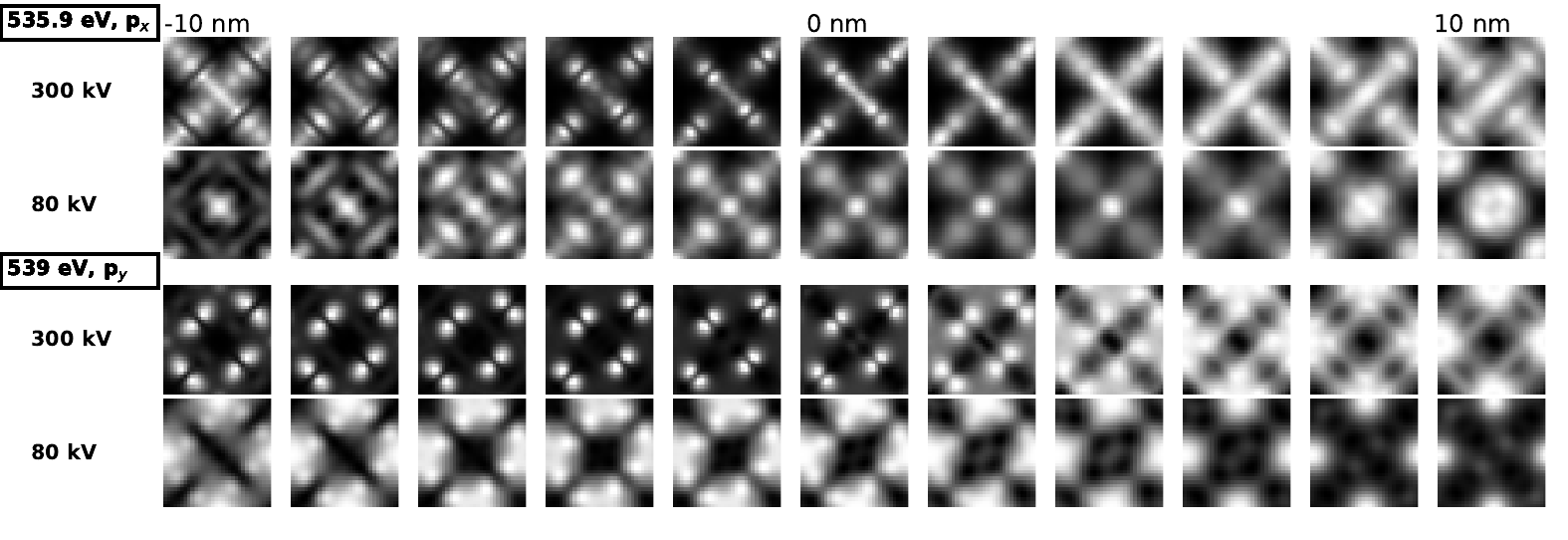}
    \caption{Simulated EFTEM images of a single rutile unit cell for varying sample thicknesses and defoci. 
   }
    \label{fig_TEM_rutile_defocus_O}
\end{figure*}

\renewcommand{\thefigure}{S10}
\begin{figure*}
    \centering
    \includegraphics{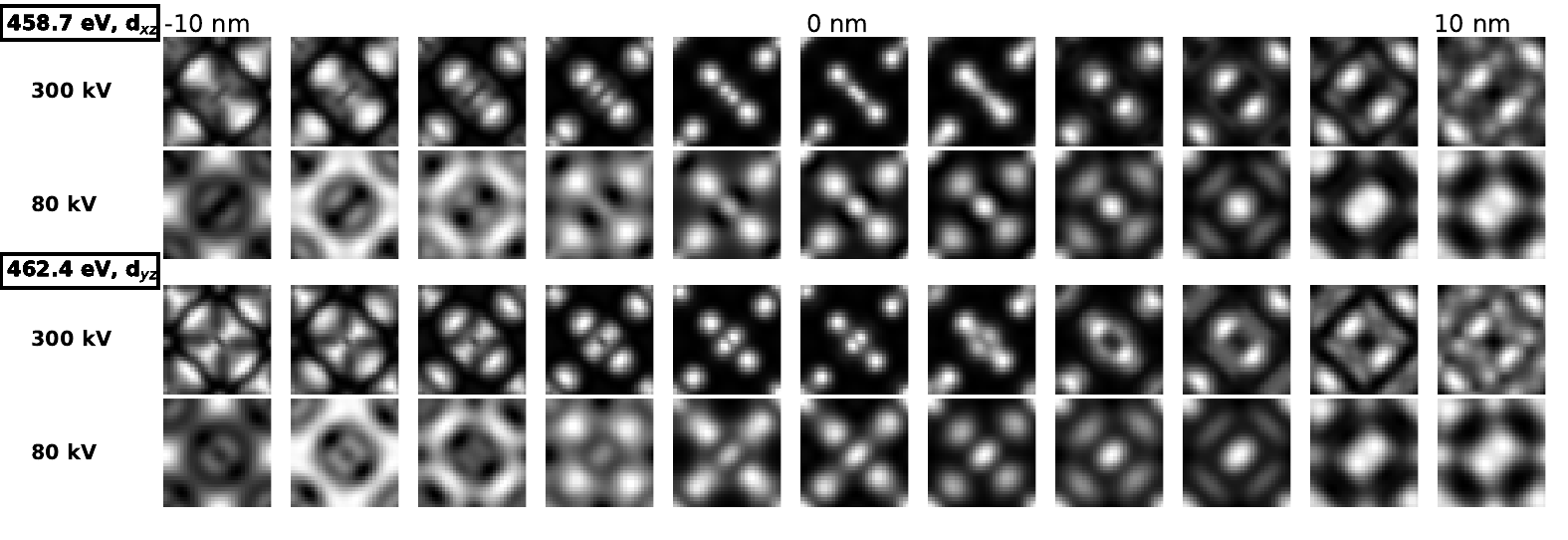}
    \caption{Simulated EFTEM images of a single rutile unit cell for varying sample thicknesses and defoci. 
   }
    \label{fig_TEM_rutile_defocus_Ti}
\end{figure*}

\renewcommand{\thefigure}{S11}
\begin{figure*}
    \centering
    \includegraphics{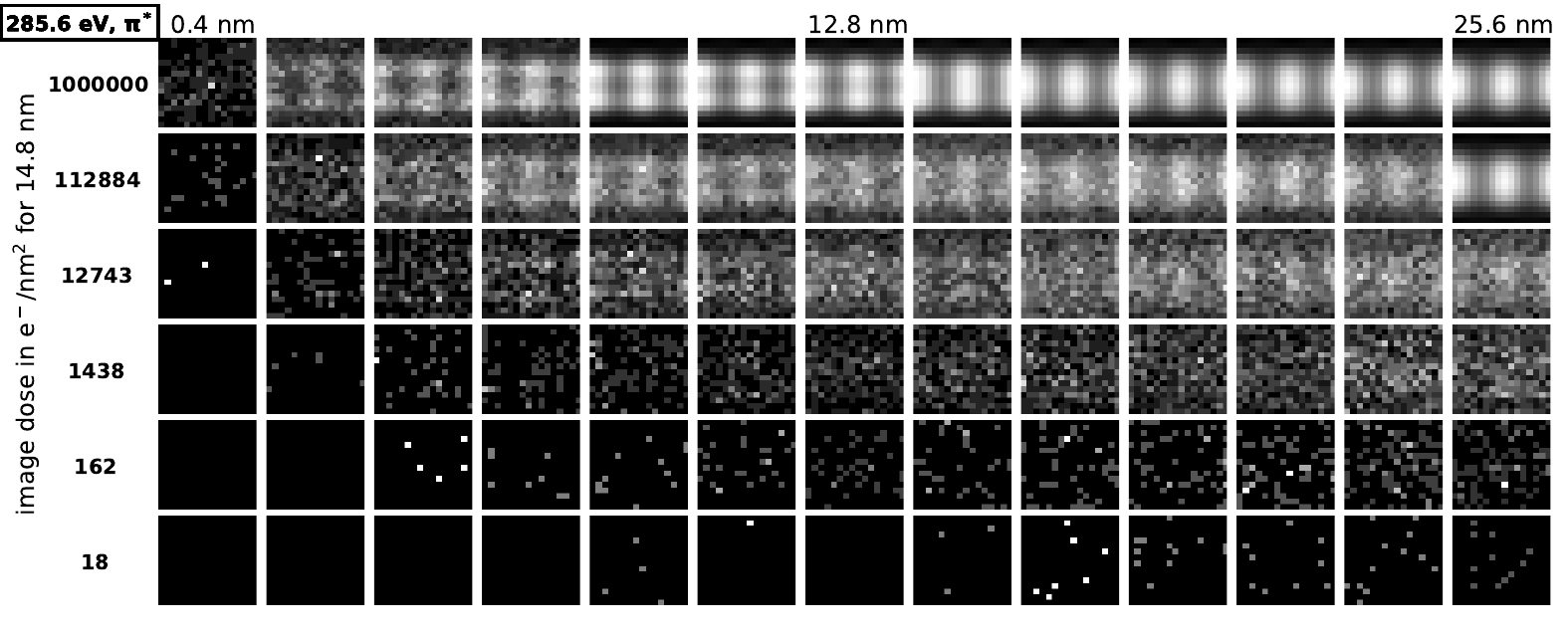}
    \caption{Simulated STEM-EELS maps of a single graphite unit cell with added shot noise for varying sample thickness.
    The image dose values are given for the 12.8~nm thick sample.
    For the other cases, the image dose is scaled with the norm of the specific infinite dose image in order to simulate a constant incident dose.
}
    \label{fig_STEM_graphite_noisewithpix}
\end{figure*}

\renewcommand{\thefigure}{S12}
\begin{figure*}
    \centering
    \includegraphics{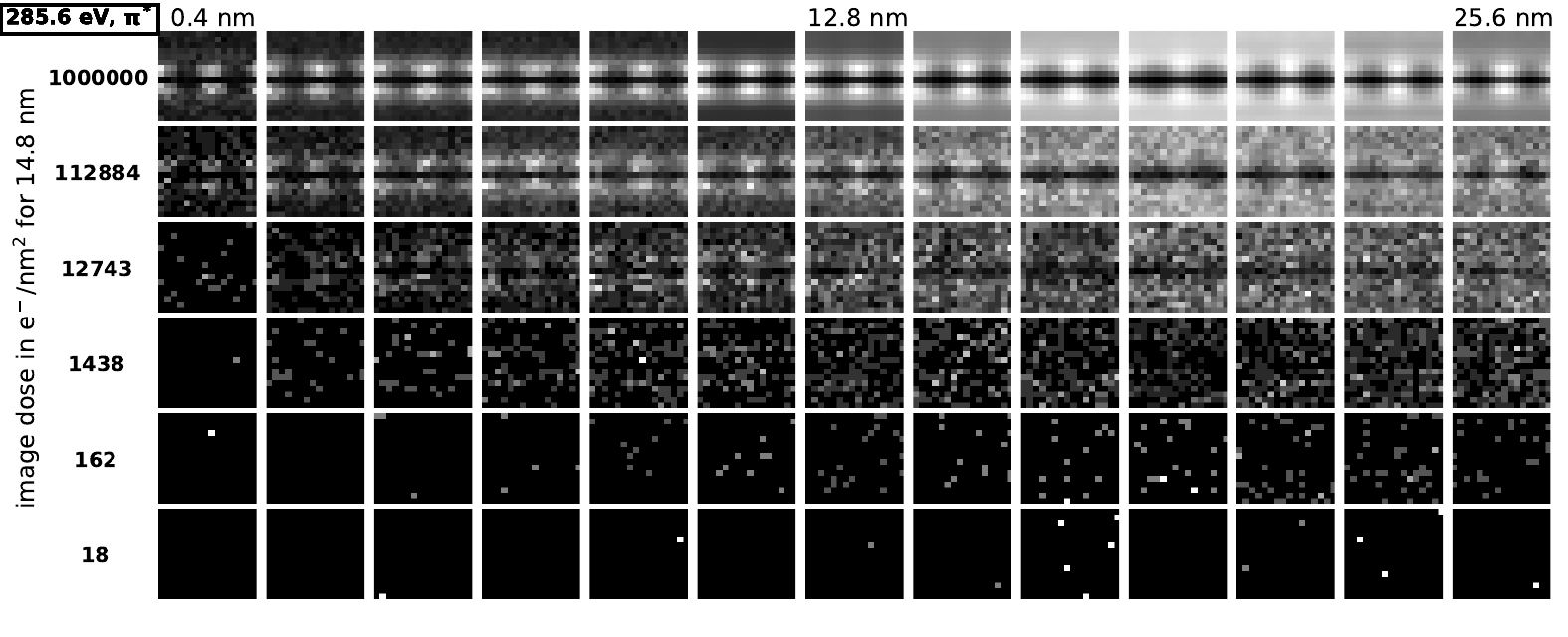}
    \caption{Simulated EFTEM images of a single graphite unit cell with added shot noise for varying sample thickness.
    The image dose values are given for the 12.8~nm thick sample.
    For the other cases, the image dose is scaled with the norm of the specific infinite dose image in order to simulate a constant incident dose.
}
    \label{fig_TEM_graphite_noisewithpix}
\end{figure*}

\renewcommand{\thefigure}{S13}
\begin{figure*}
    \centering
    \includegraphics{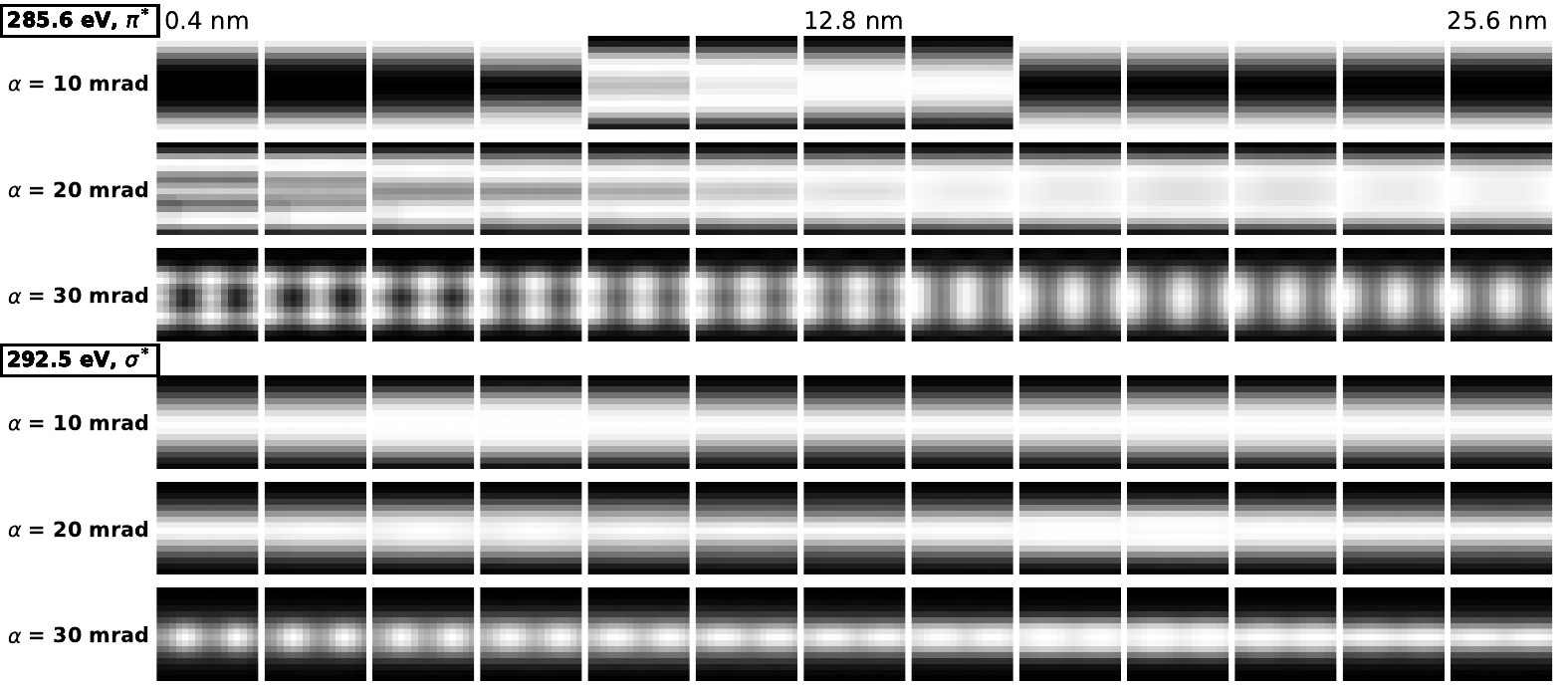}
    \caption{Simulated STEM-EELS maps of a single graphite unit cell for varying sample thicknesses and convergence semi-angles $\alpha$. 
   A collection semi-angle $\beta$ of 50~mrad was used for all images.
}
    \label{fig_STEM_graphite_convanglewithpix}
\end{figure*}

\renewcommand{\thefigure}{S14}
\begin{figure*}
    \centering
    \includegraphics{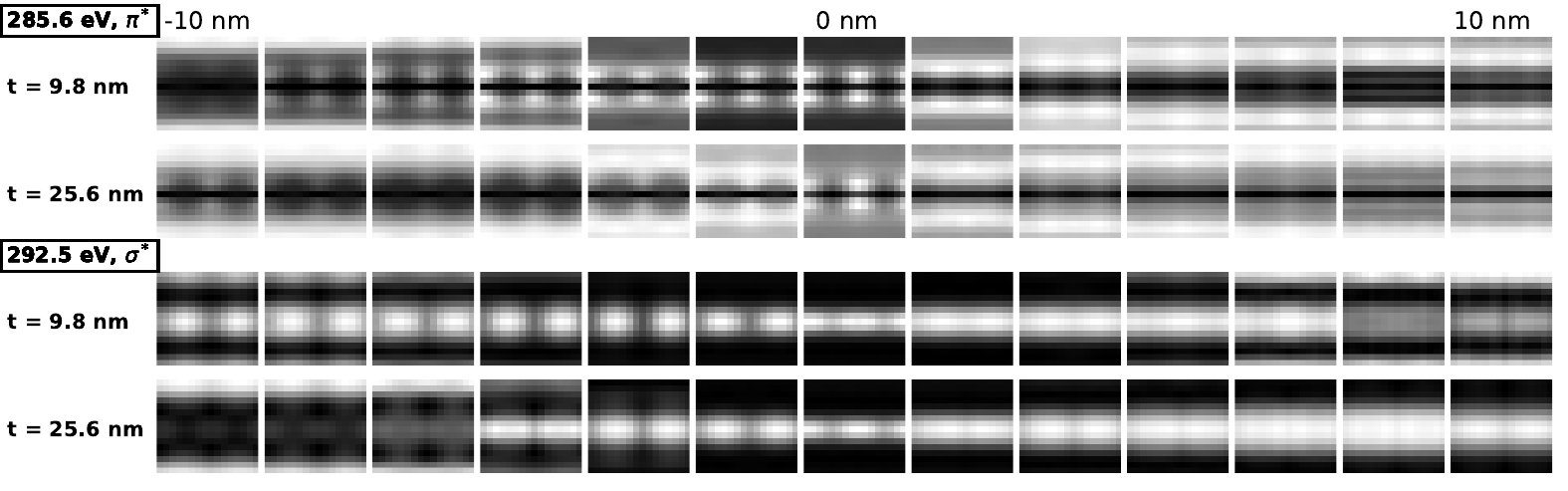}
    \caption{Simulated EFTEM images of a single graphite unit cell for varying sample thicknesses and defoci. 
   }
    \label{fig_TEM_graphite_defocus}
\end{figure*}

\renewcommand{\thefigure}{S15}
\begin{figure*}
    \centering
    \includegraphics{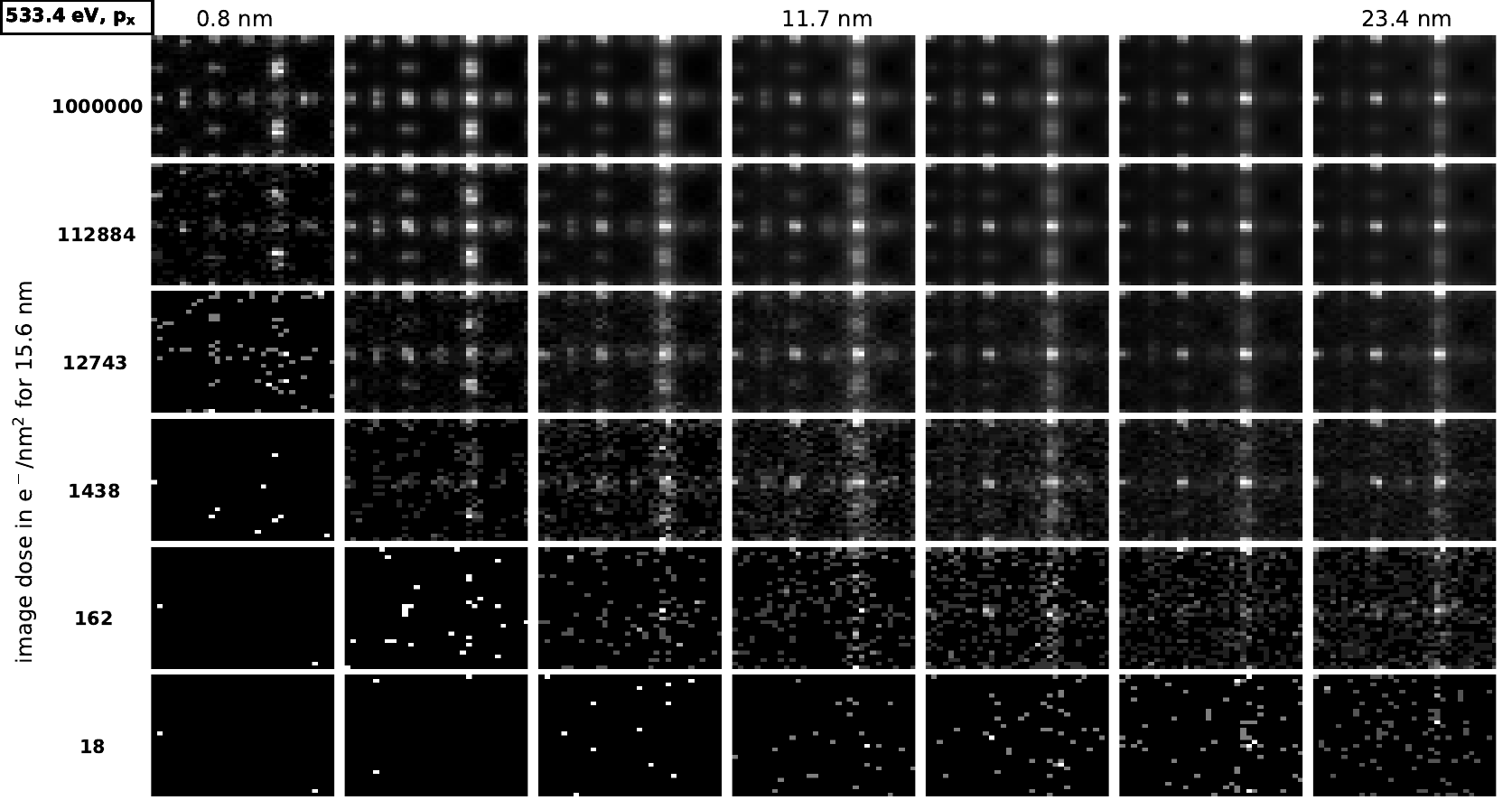}
    \caption{Simulated STEM-EELS maps of STO-LMO interface with added shot noise for varying sample thickness.
    The image dose values are given for the 16~nm thick sample.
    For the other cases, the image dose is scaled with the norm of the specific infinite dose image in order to simulate a constant incident dose.
}
    \label{fig_STEM_STOLMO_noisewithpix}
\end{figure*}

\renewcommand{\thefigure}{S16}
\begin{figure*}
    \centering
    \includegraphics{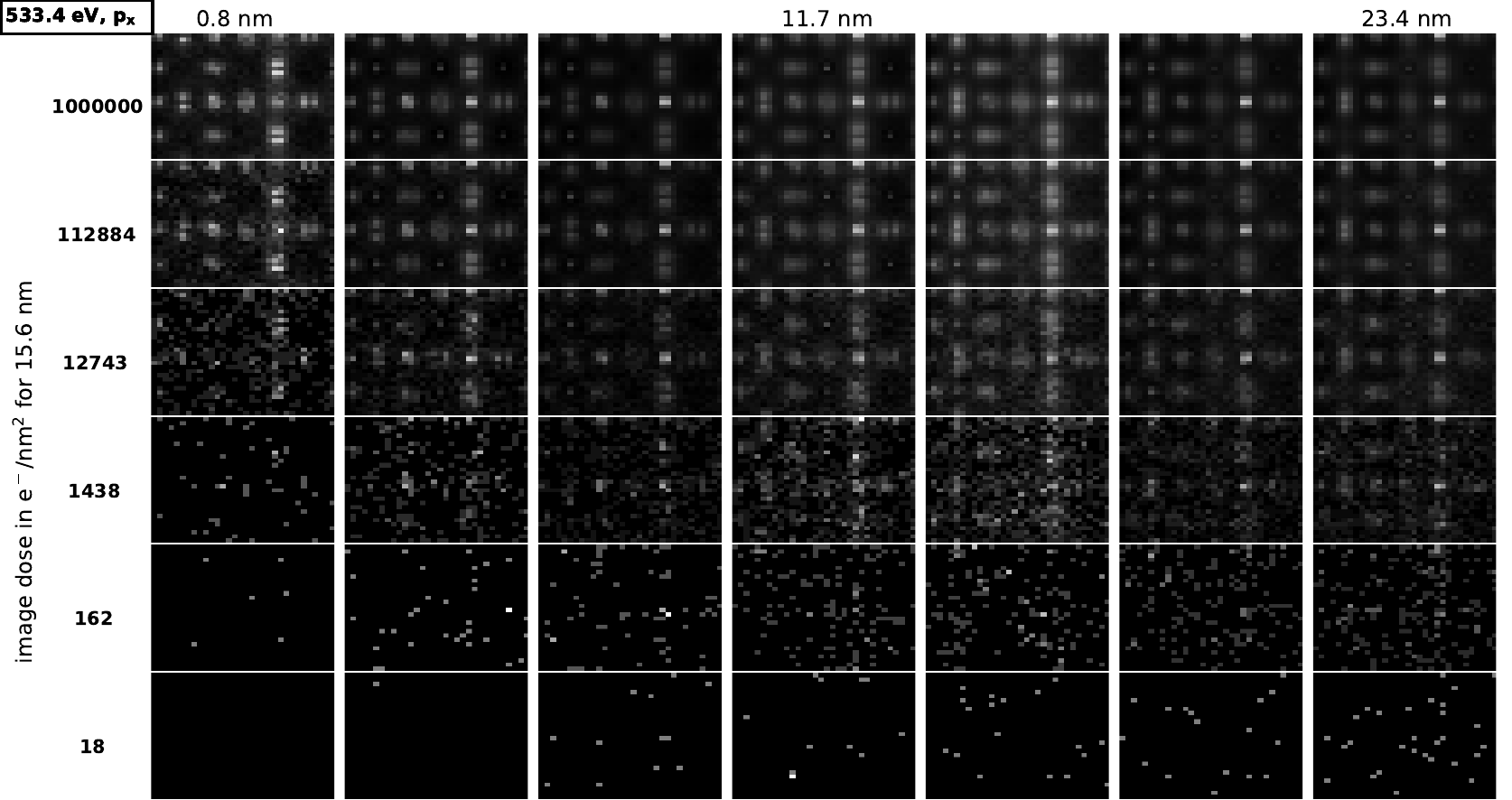}
    \caption{Simulated EFTEM images of STO-LMO interface with added shot noise for varying sample thickness.
    The image dose values are given for the 16~nm thick sample.
    For the other cases, the image dose is scaled with the norm of the specific infinite dose image in order to simulate a constant incident dose.
}
    \label{fig_TEM_STOLMO_noisewithpix}
\end{figure*}

\renewcommand{\thefigure}{S17}
\begin{figure*}
    \centering
    \includegraphics{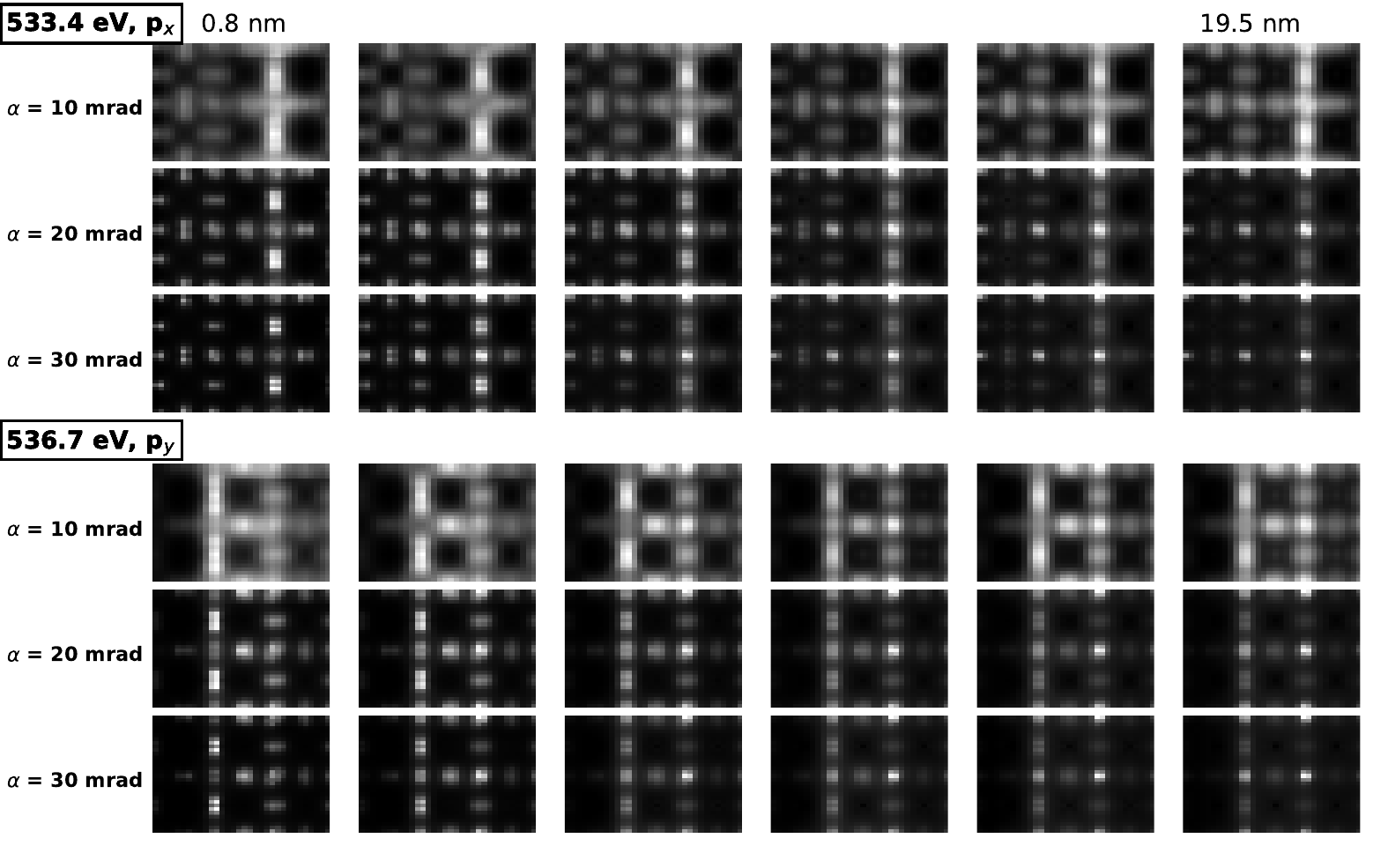}
    \caption{Simulated STEM-EELS maps of the STO-LMO interface for varying sample thicknesses and convergence semi-angles $\alpha$. 
   A collection semi-angle $\beta$ of 50~mrad was used for all images.
}
    \label{fig_STEM_STOLMO_convanglewithpix}
\end{figure*}

\renewcommand{\thefigure}{S18}
\begin{figure*}
    \centering
    \includegraphics{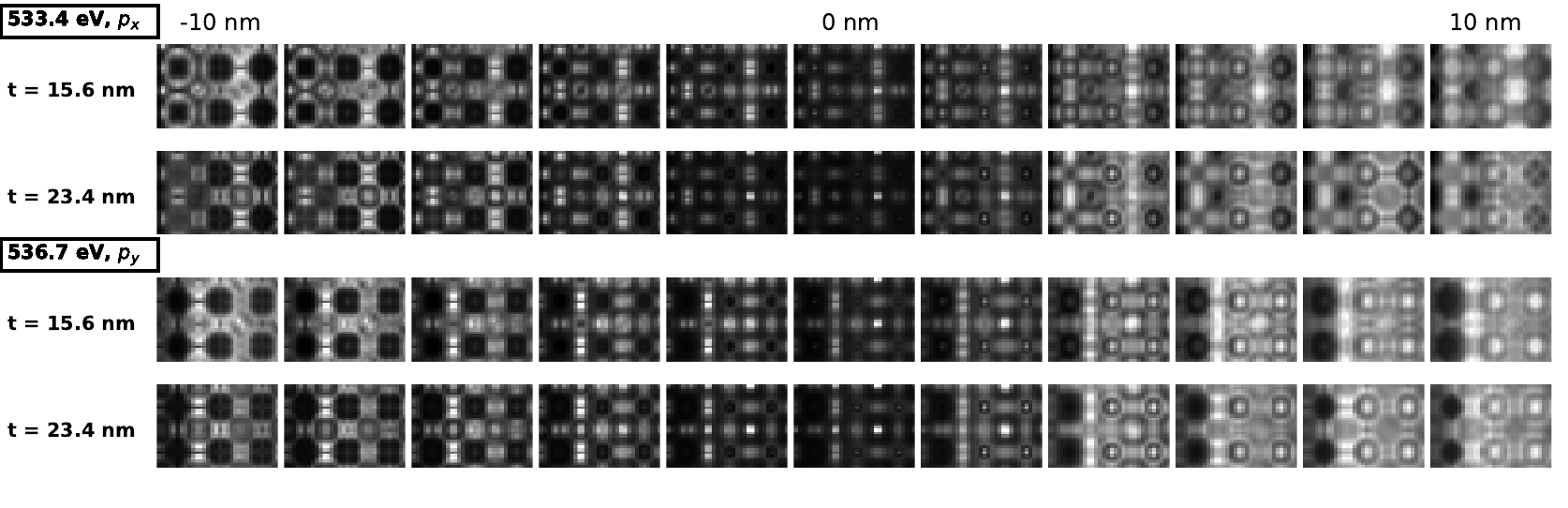}
    \caption{Simulated EFTEM images of the STO-LMO interface for varying sample thicknesses and defoci. 
   }
    \label{fig_TEM_STOLMO_defocus}
\end{figure*}

%% file: introduction.tex
\section{Introduction}
The wheel of technological progress is ever turning. 
Improvements to aberration correctors, electron detectors and spectrometers allow us to push spatial and energy resolution to continually smaller boundaries. 
Atomic resolution is now routinely achieved for both fixed-beam and scanning transmission electron microscopy (STEM) for many years \cite{Smith2010}.
Even coupled with electron-energy loss spectroscopy (EELS) \cite{Varela2009, Lazar2010, Bosman2007} or energy dispersive x-ray (EDX) analysis \cite{Kothleitner2014, Allen2012, Chu2010}, atomic resolution is now possible in the resulting energy-filtered maps. 
With orbital mapping \cite{Loeffler2017, Pardini2016, Bugnet2022}, however, we strive to fly even closer to the sun.
Contrary to elemental mapping, a tiny energy window of often less than 2~eV is necessary. 
This ensures that not the whole ionization edge is used for mapping, but only a single peak or feature of the energy loss near edge structure (ELNES), thus allowing us to map an electronic transition between specific orbitals in the sample material.
In many cases, this procedure is equivalent to imaging the spatial distribution of originally unoccupied electronic orbitals near Fermi level.
Thus, using this method allows to get a peak at the individual electronic orbitals directly responsible for basically all electronic and magnetic properties and chemical binding of a given material.
Up to recently, this was only possible for states located at the material surface through scanning tunneling microscopy \cite{Repp2006, Gross2009, Altman2015}.
Using a TEM, however, orbital mapping of bulk materials becomes possible. 
In principle, this allows the investigation of any region of interest in the material such as defects or internal boundaries between different phases or materials.
However, the demanding needs for high spatial and spectral resolution coupled with the extremely low signal-to-noise ratio due to the tiny energy windows limit the broad applicability of this promising new method for now.
Thus, it is of paramount importance to overcome these limitations by optimizing all other, more freely chooseable experimental imaging parameters such as sample thickness, acceleration voltage and electron incidence dose.\\
Aided by density functional theory and electron channelling simulations, our goal in this work is to find the perfect set of experimental parameters for three material classes: transition metal oxides (by the example of rutile), materials made up of light elements (by the example of graphite), and interfaces (by the example of the \ce{SrTiO3}--\ce{LaMnO3} heterostructure).
If there exists no perfect set we will settle for parameter ranges that produce still acceptable orbital maps.
Acceptability is gauged as objectively as possible by the use of an image difference metric and the resulting difference relative to a perfect reference image \cite{Silverstein1996, wang2004, pedersen2009}.
We have chosen rutile and graphite as their electronic structure makes these materials particularly suitable for orbital mapping while their crystal structure is still relatively simple. 
Further, they provide the opportunity to study the effects of elastic scattering on the resulting orbital maps, as graphite consists only of light atoms while rutile consists of both light and relatively heavy atoms.
The \ce{SrTiO3}--\ce{LaMnO3} heterostructure, in particular the region at/around the interface, has been chosen due to its, in comparison, complex structure and as a potential application.
Mapping core loss transitions in the vicinity of material defects and interfaces allows for potentially unrivaled insight into the local electronic and magnetic properties of the material \cite{Pardini2016}.\\
While up to now, experimental orbital maps have only been reported using scanning TEM, using EFTEM should be equally possible \cite{Loeffler2013}.
Thus, we will simulate images both for a scanning probe as well as for parallel illumination.
This allows us to compare acceptable parameter ranges between STEM and energy-filtered TEM (EFTEM) for orbital mapping whenever a direct comparison is possible.\\

%% file: methods.tex
\section{Methods} 
We calculate theoretical orbital maps by employing the multislice algorithm~\cite{Kirkland2010, Cowley1957} for the elastic propagation of the probe electron through the sample material.
The inelastic interaction between probe electrons and sample electrons is calculated with the mixed dynamic form factor~\cite{Loeffler2013, Loeffler2013phd, Schattschneider1986} based on density functional theory data obtained with WIEN2k~\cite{Blaha2020}.
Energy filtering and subsequent mapping allows us to image the corresponding transition in real space~\cite{Loeffler2017, Pardini2016}. 
The energy window will be chosen infinitely small for this work in order to decrease the numerical work load and will consist only of a single energy point.
Further, we focus only on core loss transitions, in particular on K and L edges, and interpret the resulting image as the spatial representation of the originally unoccupied orbital in the conduction band. 
EFTEM images are simulated by assuming an incidental plane wave.
STEM spectrum images, however, require an entire multislice calculation for each pixel with the convergent electron beam centered at the pixel's position.
Thus, even when taking parallel execution into account, the STEM simulations result in often restrictively long computation times.\\
For each material, we perform simulations for a representative range of sample thicknesses.
The thinnest samples are only a single unit cell thick and result, for infinite electron dose, in the qualitatively best image.
For relatively thick samples (60 -- 100 unit cells), we are unfortunately quickly limited by the aforementioned STEM computation times, which is also the reason for the varying thickest samples for different materials.
Nonetheless, for the thickness--dose investigations, we have successfully managed to include selected simulations of samples thicker than 20~nm for all materials.\\
We perform these thickness--dose investigations by including shot noise in the resulting images.
By keeping a constant ratio between the electron counts in the image and the thickness dependent signal intensity, we effectively simulate a certain electron incident dose.
Thus, without knowing the relation between the absolute electron incident dose and the resulting image dose, we can still give the relative incident dose and calculate the specific electron counts from a given reference case.
This allows us to provide a more realistic comparison between different sample thicknesses than it would be the case for infinite dose.\\
In order to automate and objectify the evaluation of the image for a specific parameter set, we employ an  image difference metric suitable for TEM~\cite{Ederer2022} 
based on the scale invariant feature transform (SIFT) algorithm~\cite{lowe1999}.
This method detects features in a reference image and compares their orientation and image gradient to the features at the same positions in any given image of interest.
Based on these differences, a total image difference is assigned to the image pair. 
An image difference of 0 hereby indicates identical images and 1 the maximal achievable difference to the reference image.
The parameters for the respective reference images are given in Tab.~\ref{table_reference_image}.

\begin{table*}
\begin{center}
\caption{Parameters of the simulated reference images.}
%\begin{tabular}{ |>{\centering\arraybackslash}m{5cm}|>{\centering\arraybackslash}m{2cm}|>{\centering\arraybackslash}m{2cm}|>{\centering\arraybackslash}m{2cm} | }
\begin{tabular}{ |m{6cm}|>{\centering\arraybackslash}m{2cm}|>{\centering\arraybackslash}m{2cm}|>{\centering\arraybackslash}m{2cm} | }
 \hline
parameters & rutile & graphite & STO-LMO\\
 \hline
 thickness (1 unit cell) [\AA{}] & 3 & 4.3 & 7.8\\
 \hline
 pixel size x [\AA{}] & 0.14 & 0.15 & 0.36\\
 \hline
 pixel size y [\AA{}] & 0.14 & 0.21 & 0.24\\
 \hline
 acceleration voltage [kV] & 300 & 60 & 300\\
 \hline
 STEM convergence semi-angle $\alpha$ [mrad] & 30 & 30 & 30\\
 \hline 
 STEM collection semi-angle $\beta$ [mrad] & 50 & 50 & 50\\
 \hline
\end{tabular}
\end{center}

\label{table_reference_image}
\end{table*}
%Further, the expected orbital shapes are simple, e.g. dumbbell shape, and with a well defined orientation in the crystal lattice which is advantageous for feature detection and comparison.
\noindent
The method is especially well suited for orbital mapping, 
as the optimal image can easily be calculated for infinite electron dose and a projected sample thickness of a single unit cell.
For STEM, we have chosen $\alpha = 30$~mrad and $\beta = 50$~mrad, as these values result in highly detailed orbital maps while still being reasonably achievable in the experiment. 
Similarly, step sizes have been chosen to allow for enough pixels per feature (orbital), so that image difference detection still performs reliably, while not too far away from typical experimental values.
Additionally, step sizes in x and y direction are different (except for rutile).
We have chosen them with the restriction that the resulting pixels per unit cell are an integer number in order to avoid artifacts.
For all simulated images, spherical aberration and incoherent source size broadening were neglected.\\
We have chosen two thresholds to further categorize the resulting image difference number, based on investigations of nano particle detectability using the SIFT metric \cite{Ederer2022}. 
Below a value of 0.18, the images that are compared can for all practical purposes be considered as identical for current state-of-the-art TEMs and the parameter set resulting in the image as optimal.
Images with difference values up to 0.3, while exhibiting already significant deviations to the reference image, still contain enough of the underlying orbital information for the parameter set to be considered acceptable here.
Accordingly, we will mark these thresholds in the relevant figures throughout this work with dark green and olive coloured contour lines or backgrounds.
We begin with an investigation of the relevant parameters for STEM-EELS and EFTEM separately.
Afterwards, we compare the methods if possible and evaluate their respective theoretical capability of achieving orbital mapping.

\section{Materials and Results}
\label{sec_matandres}
%Three different systems were investigated for this work.
%For the first material we chose rutile (\ce{TiO2}) as it is a well suited model system for orbital mapping due to its simplicity and availability. 
%We focus on the influence of channeling on the orbital map due to the relatively heavy titanium atoms.
%Secondly, Graphite as an example for a well-studied material consisting entirely of light elements.
%The final material of our investigation consists of a \ce{SrTiO3}-\ce{LaMnO3} heterostructure with the aim of imaging the influence of an interface on orbital mapping.

%% file: rutile_STEM.tex
\subsection{Rutile}
\label{sec_rutile}

The rutile phase of \ce{TiO2} is a prime target for orbital mapping due to its crystal field induced splitting of its unoccupied states near the Fermi level into $t_{2g}$-like and $e_{g}$-like bands~\cite{Grunes1982, Jiang2003, Hossain2008, Landmann2012}.
A large enough energetic separation of final states in the conduction band is crucial for orbital mapping as otherwise individual transitions to final states with different angular momentum character overlap in the resulting image.
Subsequently, all directional information of the orbital map would be lost according to Unsöld's theorem~\cite{tinkham2003}.
In the past, the t$_{2g}$ transitions of \ce{Ti} have already been successfully mapped in a proof of principle work of Löffler et al.~\cite{Loeffler2017} and more recently reproduced by Oberaigner et al.~\cite{Oberaigner2021}.\\
%oxygen
The \ce{O} K-edge yields two different relevant orbital maps, both with a characteristic dumbbell shape for the chosen projection along the \hkl[0 0 1] zone axis, due to the underlying electronic transition to a final state with p angular momentum character~\cite{Loeffler2013}. 
The resulting mapped orbitals are either aligned in the direction towards the nearest \ce{Ti} neighbour atom or perpendicular to the direction and will be defined as p$_{x}$ and p$_{y}$, respectively.
The specific energy losses of the two cases, 535.9~eV for p$_{x}$ and 539~eV for p$_{y}$, have been chosen based on the projected density of states (pDOS), ensuring high intensity of the map as well as a relatively large energy distance between the two points in order to exclude potential overlaps of experimental energy windows (see supplementary information).
Orbitals with p$_{z}$ character, in our definition parallel to the \hkl[0 0 1] zone axis and parallel to the electron beam, have little influence on the p$_{x}$ and p$_{y}$ orbital maps due to their symmetry for this particular geometry and will not be mapped individually.\\
%titanium
Compared to the \ce{O} K-edge, the \ce{Ti} L$_{2,3}$-edge is harder to interpret and has more possible final wave functions.
However, many of the cases can be disregarded for various reasons.
Due to the chosen crystallographic direction of \hkl[0 0 1], final orbitals with d$_{z^2}$ character can be excluded for symmetry reasons similar to the p$_z$ orbitals before.
d$_{x^2-y^2}$ and d$_{xy}$ exhibit a four-lobed shape in the projection only for a one unit cell thick sample but lose any resemblance to a d-orbital for samples only a few nm thick (see Fig.~S5 in the SI). 
Further, the general structure of the Ti DOS above Fermi level makes it hard to single out transitions to the d$_{x^2-y^2}$ or d$_{xy}$ orbitals, which would result in orbital maps with either a strong contribution of other final states or an unfeasibly low signal-to-noise ratio.
Thus, only d$_{xz}$ and d$_{yz}$ remain, both with dumbbell shapes resembling p orbitals for this specific projection.
The orbital at an energy loss of 458.7~eV, projection oriented towards the closest \ce{O} atoms, is defined as d$_{xz}$, while the orbital at 462.4~eV with a perpendicular projected orientation is defined as d$_{yz}$ in this work.

\subsubsection{STEM}

\begin{figure}
    \centering
    \includegraphics{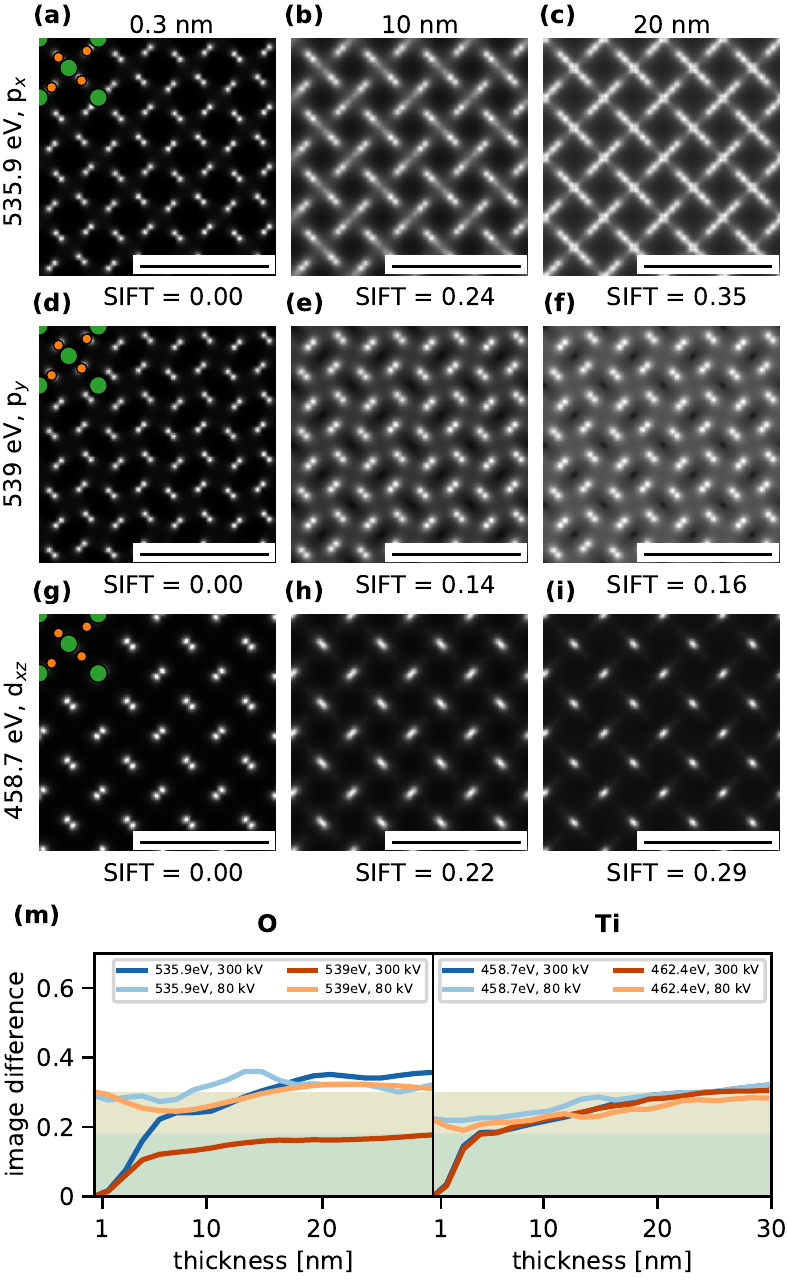}
    \caption{\textbf{(a)} - \textbf{(l)} Simulated STEM spectrum images of rutile in the \hkl[0 0 1] zone axis for the indicated sample thickness and energy loss.
    All scale bars indicate 1~nm.
    Green and orange dots mark the positions of \ce{Ti} and \ce{O} atoms, respectively.
    \textbf{(m)} SIFT image difference as function of sample thickness.
    All image differences are relative to the respective energy loss spectral image for 0.3~nm sample thickness and 300~kV acceleration voltage.
    }
    \label{fig_stem_rutile_inf}
\end{figure}

\begin{figure*}
    \centering
    \includegraphics{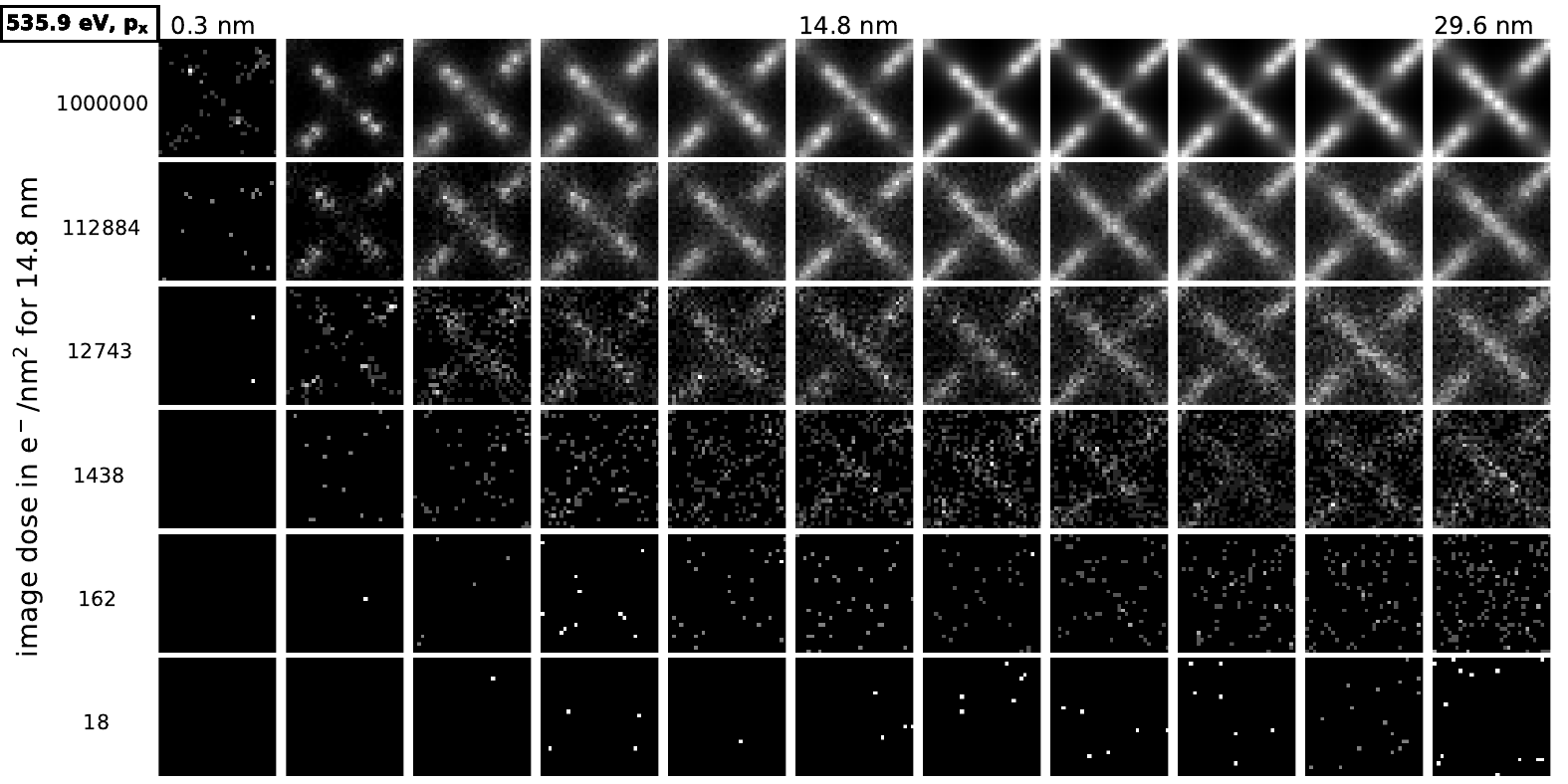}
    \caption{Simulated STEM spectrum images of a single rutile unit cell with added shot noise for varying sample thickness.
    The image dose values, chosen equidistant on a logarithmic scale, are given for the 14.8~nm thick sample.
    For the other cases, the image dose is scaled with the norm of the specific infinite dose image in order to simulate a constant incident dose.
    Shot noise is applied to a $4 \times 4$ unit cell image but only a single unit cell is shown.
    The images are individually normalized to the respective image intensity for better clarity.
}
    \label{fig_STEM_rutile_noisewithpix}
\end{figure*}

\begin{figure}
    \centering
    \includegraphics{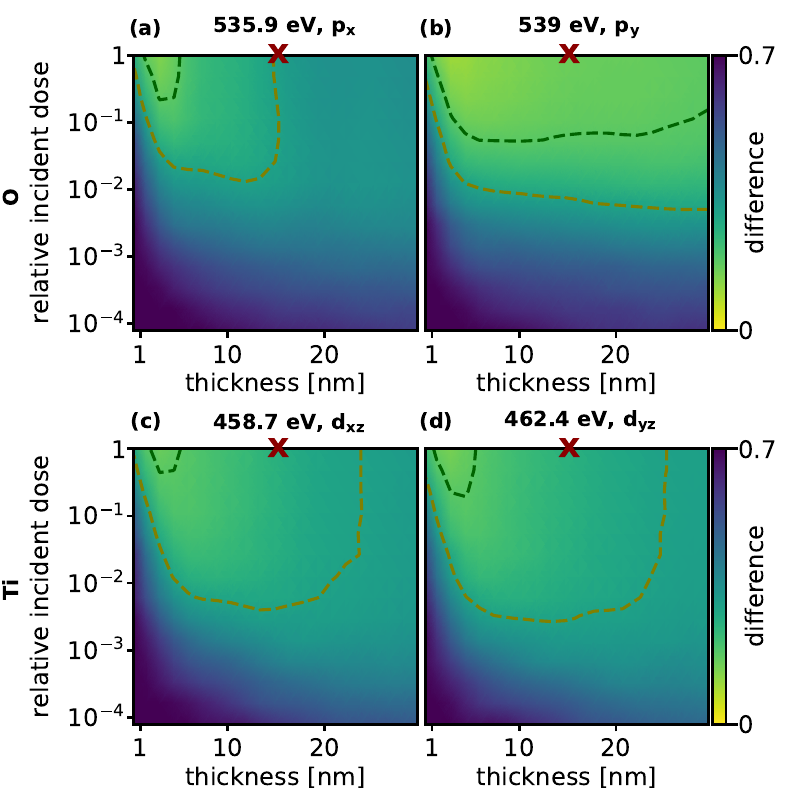}
    \caption{Image difference maps of STEM spectrum images of rutile. 
    The incident dose is measured relative to the image dose of $10^6$~e$^-$/nm$^2$ for the image belonging to a 14.8~nm thick sample, marked by the red x. 
    The reference image in all cases is calculated with infinite incident electron dose. 
    %(a), (b), (c), (d) Energy loss of 535.9 eV, 539 eV, 458.7 eV and 462.4 eV, respectively.
    The dark green and the olive contour lines indicate an image difference of 0.18 and 0.3, respectively.
}
    \label{fig_STEM_rutile_s2n}
\end{figure}

\begin{figure}
    \centering
    \includegraphics{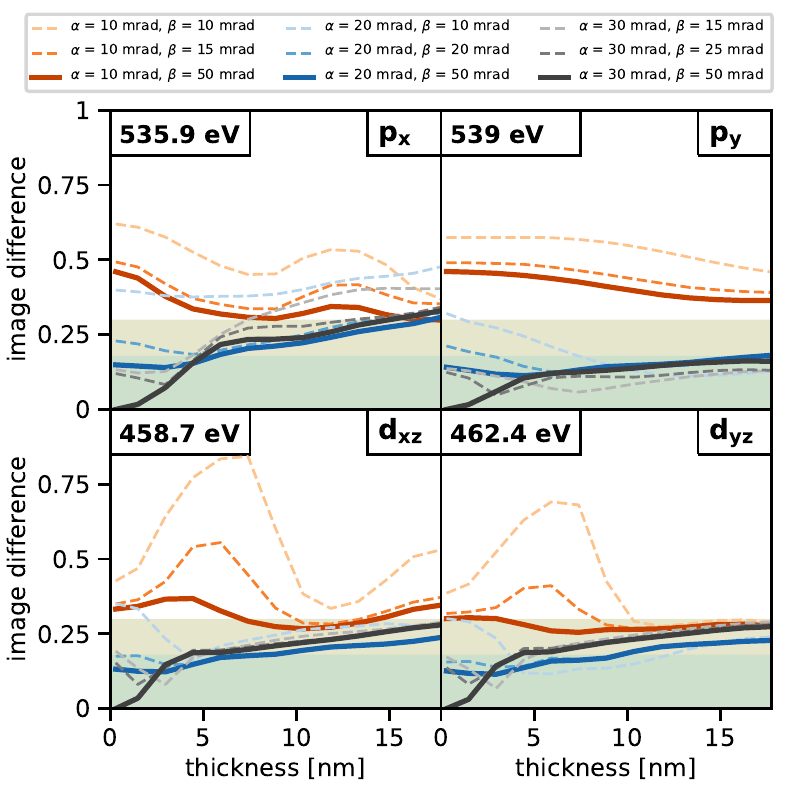}
    \caption{Rutile SIFT image difference as a function of sample thickness for an acceleration voltage of 300~kV and several convergence and collection semi-angles $\alpha$ and $\beta$.
    All image differences are relative to the orbital map for $\alpha$ = 30~mrad and $\beta$ = 50~mrad.
}
    \label{fig_STEM_rutile_conv}
\end{figure}

In the following, we present the simulated STEM spectral images of the four cases discussed in the previous section.
We focus on the influence of sample thickness, incident electron dose and convergence (collection) semi-angle $\alpha$ ($\beta$) on the image quality, i.e. the recognizability of the original projected orbital shapes.\\
The two upper rows in Fig.~\ref{fig_stem_rutile_inf} show the two relevant cases of the \ce{O}-K edge.
They exhibit a drastically different sample thickness dependence, due to the preferred inelastic scattering direction parallel to the orbital orientation.
In the case of the p$_x$ orbital, when the sample thickness is increased, the perfect orbital shape deteriorates in the image due to the channeling effect~\cite{Kirkland2010, Loane1988, Schattschneider2003} and due to partial waves created at neighbouring atomic columns in the bulk sample where the incident beam is already less focused.
Thus, with increasing sample thickness, increasingly more intensity from the orbital shape is transferred to the closest \ce{Ti} atom in the image, 
with the result that the two p$_x$ lobes appear as a single, long ellipse.
The channeling effect is significantly less pronounced for the p$_y$ orbital due to the preferred scattering direction in between the (distorted) octahedra of the rutile crystal structure.
Even for a projected sample thickness of 20~nm, the only perceivable change in the spectral image is a slight background intensity increase between the atoms.
The dumbbell shape is mostly preserved.
This is reflected in the comparatively low image difference indicated by the SIFT algorithm. \\
The third row in Fig.~\ref{fig_stem_rutile_inf} shows the transitions to final states with d$_{xz}$ character of the \ce{Ti}-L$_3$ edge.
Transitions to d$_{yz}$ states are not explicitly shown, as they are basically equivalent to Fig.~\ref{fig_stem_rutile_inf}(g)-(i) except for the 90$^\circ$ rotation of the orbital shapes. 
Unlike for the \ce{O} orbitals, here, the comparatively heavy \ce{Ti} atom is at the center of the orbitals.
This leads to a focusing effect of the scattered electrons and with increasing thickness, the dumbbell shape shifts to a concentrated ellipse.
We want to stress at this point that this is not a shortcoming that can be overcome by technical improvements of, e.g., the spatial/spectral resolution or 
the spot size of the electron beam, but solely by reducing the sample thickness, sometimes to unrealistic sizes.
Nevertheless, while for this energy, the dumbbell shape is lost, the directional dependence is preserved for any thickness, thus the two relevant cases 
with final orbital d$_{xz}$ and d$_{xy}$ can still be distinguished.\\
The last row of Fig.~\ref{fig_stem_rutile_inf} shows in detail the thickness dependence of the image difference of the \ce{O} and \ce{Ti} energy-filtered STEM images. 
For an acceleration voltage of 300~kV, all energy losses display a drastic difference increase in the first few nm followed by an only slight increase over the next tens of nm.
However, for a voltage of 80~kV, the difference to the 300~kV reference image either fluctuates around 0.3 or increases only slightly with sample thickness.
Interestingly enough, for nearly all cases, the choice of acceleration voltage and sample thickness (above 5~nm) seems to have little to no influence on the quality of the resulting orbital map. 
The only exception is the \ce{O} p$_y$ orbital for an acceleration voltage of 300~kV, displaying a considerably lower image difference compared to the other cases 
due to the aforementioned reasons.\\
\noindent
While the simulated thickness scans as in Fig.~\ref{fig_stem_rutile_inf} for an infinite electron dose are very useful for indicating general trends of the orbital maps in regards to the sample thickness, a more holistic view is gained when shot noise and the resulting signal-to-noise ratio is considered.
We effectively simulate a certain electron incident dose by keeping a constant ratio between the electron counts in the image and the thickness dependent signal intensity.
Thus, without knowing the relation between the absolute electron incident dose and the resulting image dose in a given microscope, we can still give the relative incident dose and calculate the specific electron counts from a given reference case.
We choose for the reference image dose 10$^6$~e$^-$/nm$^2$ for the 14.8~nm thick sample, as this should result in a representative range of image doses.
%We simulate a finite incident electron dose and a resulting signal that depends on the number of atoms and, thus, on the thickness of the material.
A broad range of relative incident doses over several magnitudes is investigated. 
The image doses range from a few electrons for the whole image to many thousands, effectively approaching an infinite dose image. 
The resulting noisy images of the p$_x$ orbitals are shown in Fig.~\ref{fig_STEM_rutile_noisewithpix}.\\
In order to be able to properly gauge the effect the diminished signal-to-noise ratio has on the image quality, we again apply the image difference metric and use for each energy loss the respective infinite dose reference image with parameters according to Table~\ref{table_reference_image}.
However, as all images are normalised for the comparison, changes in the total image intensity are not factored into the image difference.  
%while direct quantitative comparison to experimental measurements with a certain incident dose is 
%at this point not possible yet, nevertheless, a more comprehensive picture is gained.
The results for the previously discussed four relevant energy losses are presented in Fig.~\ref{fig_STEM_rutile_s2n}.
All cases exhibit the same general behaviour.
Basically independent of sample thickness, an incident dose not lower than 1~\% of the reference incident dose is required for the orbitals to be distinguishable from the noise.
Contrary to the previous cases with infinite electron dose, the minimum for a given incident electron dose is no longer found at the thickness of a single unit cell.
For very low doses, the minimum actually is at the thickest investigated sample, although we conjecture that the image difference would still drop for a 
higher sample thickness.
For very high doses, on the other hand, an actual minimum is found at about 2~nm thickness, converging toward the value for an infinite electron dose.\\
The map of \ce{O} p$_y$ orbital (539~eV) shows a slightly different thickness dependence than the other cases, in agreement with the findings from Fig.~\ref{fig_stem_rutile_inf}.
Above a thickness of about 5~nm, the image difference stays at a relatively low level and increases only negligibly with increasing thickness.
This behaviour makes it the preferred case for experimental orbital mapping as long as high enough electron incident rates can be achieved without damaging the sample.
Finally, for rutile STEM, we investigate the influence of the electron beam convergence semi-angle $\alpha$ on the quality of the orbital maps (Fig.~\ref{fig_STEM_rutile_conv}).
For a single unit cell thick sample, we find that a higher convergence angle always results in a qualitatively better orbital map.
This result is not surprising for simulations without aberrations.
As such, we have chosen the spectral images for the highest investigated $\alpha$ as our reference images. 
It appears, however, that for increasing sample thickness, the $\alpha$ dependence of the difference almost completely disappears.
Only for the smallest convergence semi-angle of 10~mrad a (marginally) higher image difference can be observed.
We can conclude from this data that for realistic sample thicknesses, the choice of convergence semi-angle has little influence on the final orbital map.
When, in addition, various collection semi-angles $\beta$ are considered, it becomes apparent that, with few exceptions, $\beta < \alpha$ leads to higher image differences.
Further, with $\beta > \alpha$ the change of image difference with increasing $\beta$ is negligible.
Considering, however, that an increase of $\beta$ leads to a higher signal intensity and, thus, better signal-to-noise ratio, a larger $\beta$ is advantageous in general, as long as large angle scattering effects like thermal diffuse scattering do not contribute to the signal.

%% file: rutile_TEM.tex
\subsubsection{TEM}
\begin{figure}
    \centering
    \includegraphics{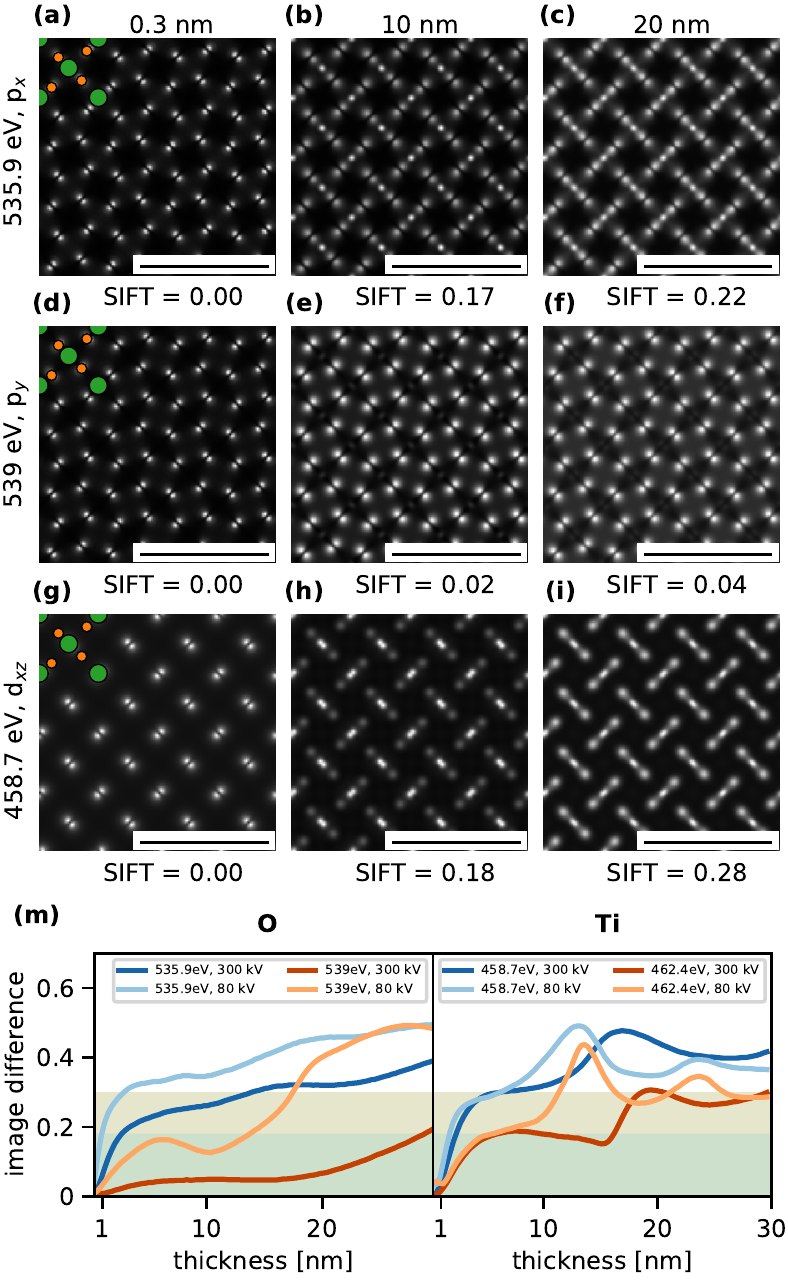}
    \caption{(a) - (l) Simulated energy-filtered TEM images of rutile in the \hkl[0 0 1] zone axis for the indicated sample thickness and energy loss.
    All scale bars indicate 1~nm. 
    Green and orange dots mark the positions of \ce{Ti} and \ce{O} atoms, respectively.
    (m) SIFT image difference as function of sample thickness.
    All image differences are relative to the respective energy-filtered image for 0.3~nm sample thickness and 300~kV acceleration voltage.  
    }
    \label{fig_TEM_rutile_inf}
\end{figure}

\begin{figure}
    \centering
    \includegraphics{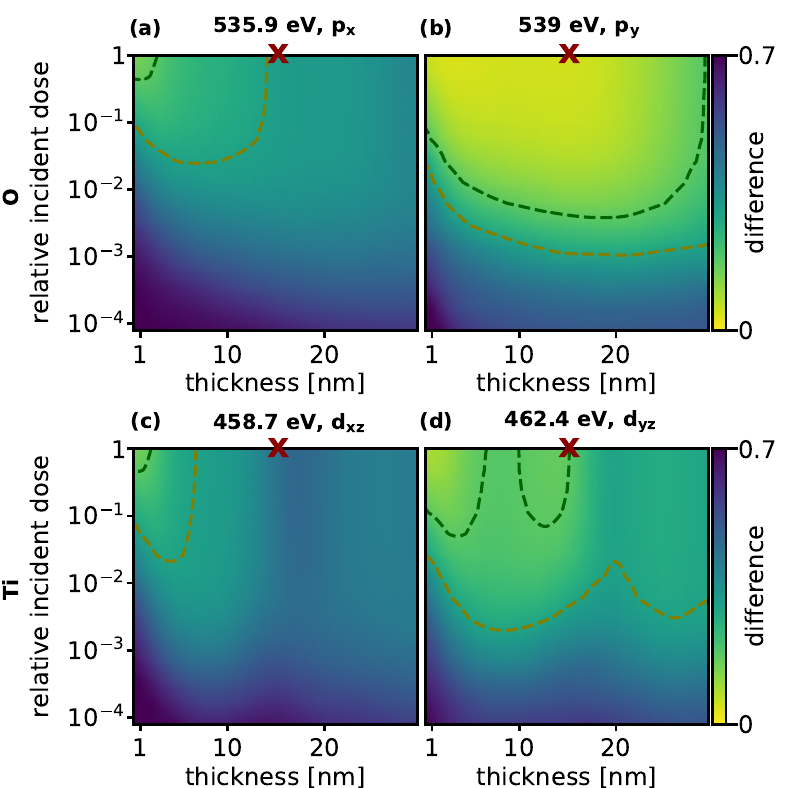}
    \caption{Image difference maps of EFTEM images of rutile. 
    The incident dose is measured relative to the image dose of $10^6$~e$^-$/nm$^2$ for the image belonging to a 14.8~nm thick sample, marked by the red x. 
    The reference image in all cases is calculated with infinite incident electron dose.
    %(a), (b), (c), (d) Energy loss of 535.9 eV, 539 eV, 458.7 eV and 462.4 eV, respectively.
    The dark green and the olive contour lines indicate an image difference of 0.18 and 0.3, respectively.
}
    \label{fig_TEM_rutile_s2n}
\end{figure}

\begin{figure}
    \centering
    \includegraphics{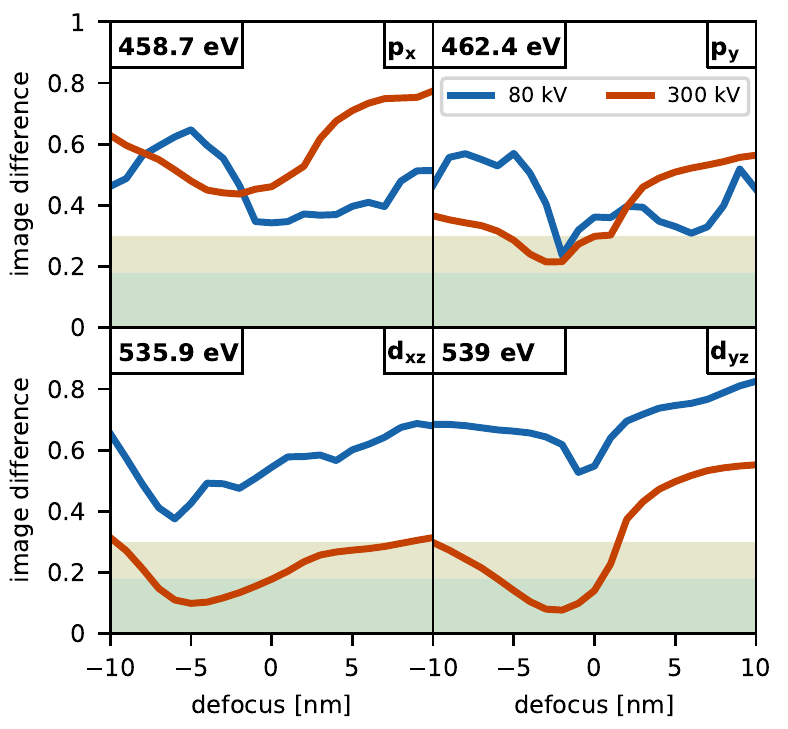}
    \caption{SIFT image difference of EFTEM images as a function of defocus for acceleration voltages of 80~kV and 300~kV and a sample thickness of 29.6~nm.
    All image differences are relative to the orbital map for 0~defocus.
}
    \label{fig_TEM_rutile_defocus}
\end{figure}

In the following, we present the simulated energy-filtered TEM (EFTEM) images for the four energy losses already discussed previously.
For simulation parameters where comparison to STEM is possible, such as sample thickness and incident electron dose, we will proceed fully analogously as in the previous section and compare the methods from a theoretical point of view.
Instead of convergence angles, however, we will focus on the effect of (a small) defocus on the quality of the EFTEM orbital map.\\
The upper two rows in Fig.~\ref{fig_TEM_rutile_inf} show very similar thickness dependence of the \ce{O} EFTEM maps compared to the STEM spectrum images.
However, for both energy losses, the degradation of the orbital shape due to elastic channeling seems even less pronounced. 
Nevertheless, the transition to the p$_y$ orbital, situated at 539~eV energy loss, still appears to be the most favourable of the four discussed transitions for orbital mapping.
While the \ce{O} orbitals display less elastic channeling compared to STEM, the \ce{Ti} orbitals display the opposite behaviour. 
Both cases show a strong transfer of intensity from the orbital to the \ce{O} atoms, fluctuating with the sample thickness according to the Pendellösung.
Further, it appears that the specific orientation of the orbital does not influence the amount of elastic channeling that occurs.\\
The thickness dependent image difference further highlights the difference in thickness dependence between EFTEM  [Fig.~\ref{fig_TEM_rutile_inf}(m)] and STEM [Fig.~\ref{fig_stem_rutile_inf}(m)] images.
Firstly, with the exception of the \ce{O} p$_y$ orbital, image differences are generally higher for EFTEM. 
At a thickness of one unit cell, however, the difference between 80 kV and 300 kV acceleration voltage are almost non-existent for EFTEM (under ideal conditions), in line with the reciprocity theorem \cite{Laue1948, Cowley1969}.
Secondly,  the image difference fluctuates with sample thickness, especially noticeable for the \ce{Ti} orbitals.
While for STEM, a higher sample thickness generally means a higher image difference, this is not the case for EFTEM.
On the contrary, in multiple cases, the image quality actually improves with thickness in some regions due to Pendellösung effects.
This shows that finding the optimal sample thickness for EFTEM orbital mapping is entirely dependent on the particular energy loss, acceleration voltage, sample, and orientation.\\
\noindent
Similar to the previous section, the dose dependent image difference maps (Fig.~\ref{fig_TEM_rutile_s2n}) further support the findings from the images with infinite electron dose. 
In general, the image difference maps for EFTEM are very similar to the STEM image difference maps (Fig.~\ref{fig_STEM_rutile_s2n}). 
Notable exceptions are the even less pronounced thickness dependent changes for \ce{O} p$_{y}$ images and the image difference fluctuations with sample thickness for \ce{Ti} d$_{xz}$ and d$_{yz}$.\\
Finally, we investigate the influence of defocus on the EFTEM maps for the largest rutile sample thickness investigated in this work, i.e., a sample 100~unit cells thick, corresponding to 29.6~nm (Fig.~\ref{fig_TEM_rutile_defocus}).
We use the following sign convention: a negative defocus shifts the objective focus point from the bottom of the sample higher, i.e., inside the sample.
For 80~kV, it becomes apparent that the image difference introduced by the sample thickness is too large for the defocus to be able to positively influence it to an acceptable range, with the singular exception of $-3$~nm defocus for the p$_y$ orbital.
Further, no clear trend can be discerned concerning the influence a defocus has on the image difference.
For 300~kV, however, a clear trend is visible.
A slightly negative defocus actually improves the similarity between the EFTEM image and the reference image, with the image difference minimum between $-3$~nm and $-5$~nm for all investigated energy losses. 
A positive defocus, however, has the opposite influence and very quickly ($>2$~nm) worsens the energy-filtered image to the point of unrecognizability of any orbitals.

%% file: graphite.tex
\subsection{Graphite}
For our second material, we have chosen graphite, as it consists entirely of light elements, in contrast to the material of the previous section, rutile.
Graphite, especially its two-dimensional single layer form graphene, has received considerable attention in the last $\sim$20 years \cite{Katsnelson2007, Geim2007, Novoselov2004}, in particular, but not exclusively, due to its potential manifold applications in electronics.\\
Graphite consists of carbon layers in a hexagonal structure weakly bound by van der Waals forces.
The high rotational symmetry of the individual graphene layers dictates the crystallograhic direction necessary for orbital mapping. 
Conventionally, for graphene and similar two-dimensional materials, a top-down view is employed.
In our case, however, this direction would result only in rotationally symmetric maps with negligible differences between different energy loss windows for pristine graphite \cite{Pardini2016, Tavabi2011}.
Thus, a side view is required and we opt to choose the \hkl[1 0 -1 0] zone axis, the same as in \cite{Bugnet2022}.\\
Two peaks dominate the DOS above the Fermi energy and, likewise, the energy loss near edge structure (ELNES) of the graphite K-edge.
The first peak is a result of the $\pi^\ast$ orbital, having mainly p$_z$ angular momentum character.
The second peak is a result of the $\sigma^\ast$ orbital with a mixture of p$_x$, p$_y$ and s atomic orbitals that combine to a sp$^2$ orbital.
We simulate energy-filtered maps at 285.6~eV and 292.5~eV energy loss for the $\pi^\ast$ and the $\sigma^\ast$ orbitals, respectively. 
Contrary to the previous case of rutile, here, the slow changing DOS would allow for relatively large energy windows where the resulting orbital maps are basically the same.
For graphite, we only consider a high tension of 60~kV as a higher acceleration voltage would already induce significant knock-on damage in the sample.~\cite{Meyer2012}
\subsubsection{STEM}
\begin{figure}
    \centering
    \includegraphics{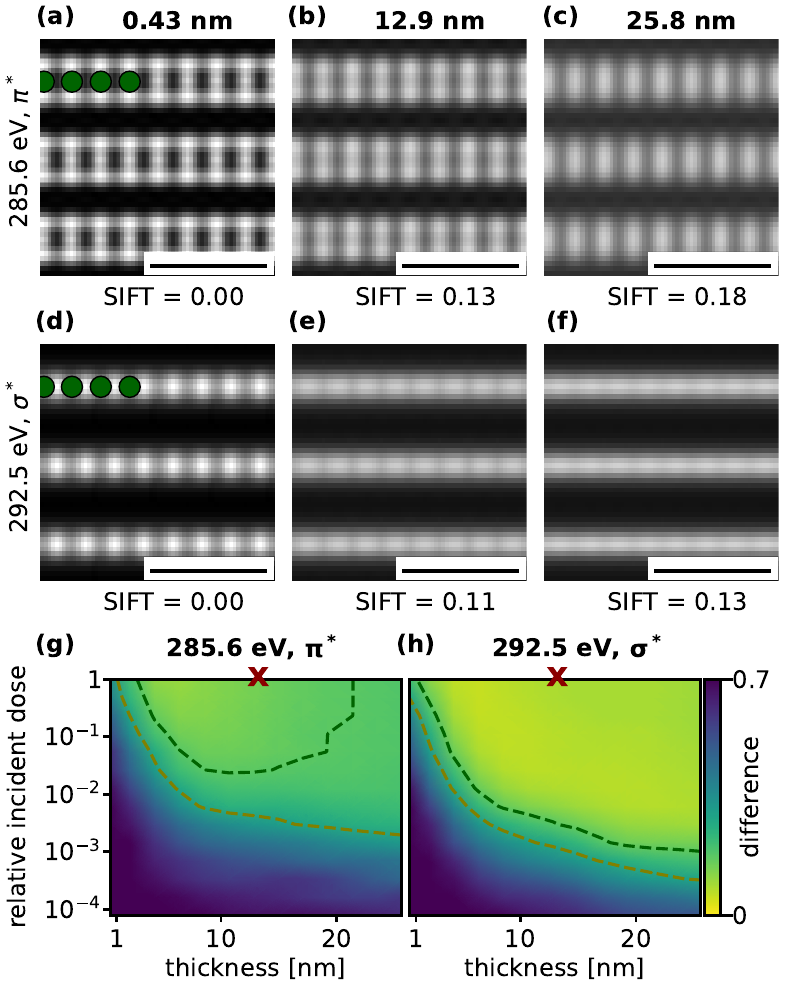}
    \caption{(a) - (f) Simulated STEM spectrum images of graphite in the \hkl[1 0 -1 0] zone axis for the indicated sample thickness and energy loss with an infinite electron dose.
    All scale bars indicate 0.5~nm. 
    Green dots mark the positions of \ce{C} atoms.
    (g), (h) Image difference maps of the graphite spectrum images for the indicated energy losses.
    The incident dose is measured relative to the image dose of $10^6$~e$^-$/nm$^2$ for the image of a 12.8~nm thick sample, marked by the red x.
    The reference image in all cases is calculated with infinite electron dose.
    The dark green and olive contour lines indicate an image difference of 0.18 and 0.3, respectively.
}
    \label{fig_STEM_graphite_s2n}
\end{figure}

\begin{figure}
    \centering
    \includegraphics{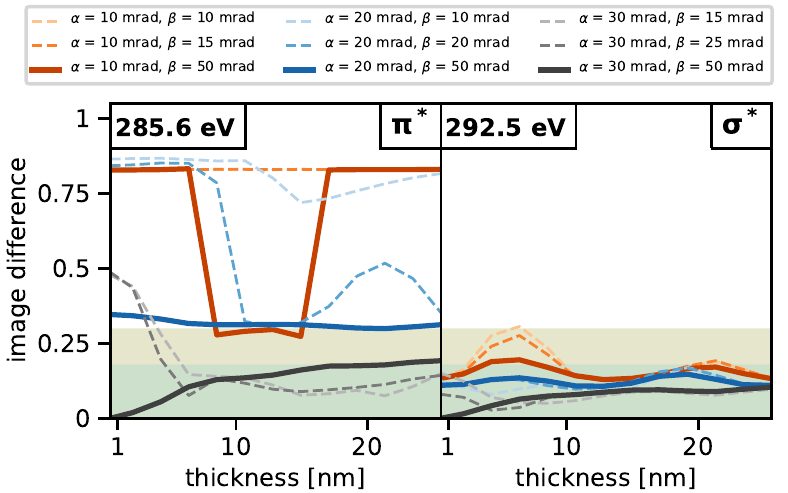}
    \caption{Graphite SIFT image difference as a function of sample thickness for an acceleration voltage of 60~kV and several convergence and collection semi-angles $\alpha$ and $\beta$. 
    All image differences are relative to the orbital map for $\alpha$ = 30~mrad and $\beta$ = 50~mrad. 
}
    \label{fig_STEM_graphite_conv}
\end{figure}

The resulting STEM spectrum images are shown in Fig.~\ref{fig_STEM_graphite_s2n}.
In the upper row, spectrum images for an energy loss of 285.6~eV are displayed. 
For a single unit cell thick sample, the characteristic dumbbell shape of the p$_z$ orbitals is clearly visible.
However, even for this thinnest possible sample, there is a small local intensity maximum at the positions of the \ce{C} atoms due to the elastic scattering caused by the atomic columns.
This effect is more pronounced with increasing sample thickness, resulting in the complete deterioration of the imaged dumbbell shapes to smeared out ellipses. 
As the sample thickness onset for this reshaping is already at thicknesses smaller than 5~nm, in practice, there is effectively no way to prevent it experimentally.
Nevertheless, due to the channeling effect also being present in the reference image, the resulting image differences are relatively small.
The electron dose dependent image difference map in Fig.~\ref{fig_STEM_graphite_s2n}(g) further confirms this finding.\\
%Above a certain incident dose and sample thickness the image difference is basically independent of the sample thickness.\\
A similar trend can be observed for the spectrum images for an energy loss of 292.5~eV. 
Due to the projection, the $\sigma^\ast$ orbitals appear as circular shapes significantly more confined to the atomic rows compared to the $\pi^\ast$ orbitals.
Horizontally, however, the intensity maxima are localized between the carbon positions.
Thus, the elastic channeling effect of a thicker sample results primarily in a horizontal blurring and a comparatively low image difference to the reference image. 
In the dose-thickness difference map [Fig.~\ref{fig_STEM_graphite_s2n}(h)], a similar trend to Fig.~\ref{fig_STEM_graphite_s2n}(g), albeit more pronounced, can be observed.
There is only little further increase of the image difference above 7~nm for all electron doses above 1~\% relative dose. 
In contrast to the difference map of the $\pi^\ast$ orbital, however, for medium to high electron doses, the image difference converges to the low value of 0.1.\\
\noindent 
In general, the graphite spectrum images appear to be a little bit more robust concerning thickness variations compared to the rutile spectrum images of the previous section. 
Possible explanations for this robustness are the low high tension of 60~kV (the reference image itself appears already less detailed) and that carbon is the only element in the unit cell (less chance for noticeable dechanneling).\\
For changes in the electron beam convergence semi-angle $\alpha$, however, there are significant differences to before.
In the case of rutile, we arrived at the more or less general conclusion that for realistic sample thicknesses, the convergence angle played no significant role at all.
For the $\pi^\ast$ orbital of graphite, however, the choice of $\alpha$ is crucial [Fig.~\ref{fig_STEM_graphite_conv}(a)].
The larger beam crossover for $\alpha = 20$~mrad results in the loss of the ability to resolve the individual dumbbells in the image. 
Further, $\alpha = 10$~mrad results in an effect similar to a contrast inversion.
The beam crossover is large enough that the probe can excite inelastic transitions in two layers when the beam is centered between layers. 
Centered on the atomic layers, however, only transitions in the respective layer can be triggered, resulting in less intensity in the STEM-EELS map and a contrast inversion.
For a sample thickness between 9~nm and 15~nm, however, we theorize that channeling effects again reverse this effect.
The result is a large decrease in image difference for this thickness range.
The thickness and $\alpha$ dependent STEM-EELS maps can be seen in Fig.~S13 in the supplementary information. 
In the case of the $\sigma^\ast$ orbital, however, the convergence angle again has very little influence on the quality of the orbital map [Fig.~\ref{fig_STEM_graphite_conv}(b)].

\noindent 
\subsubsection{TEM}
\begin{figure}
    \centering
    \includegraphics{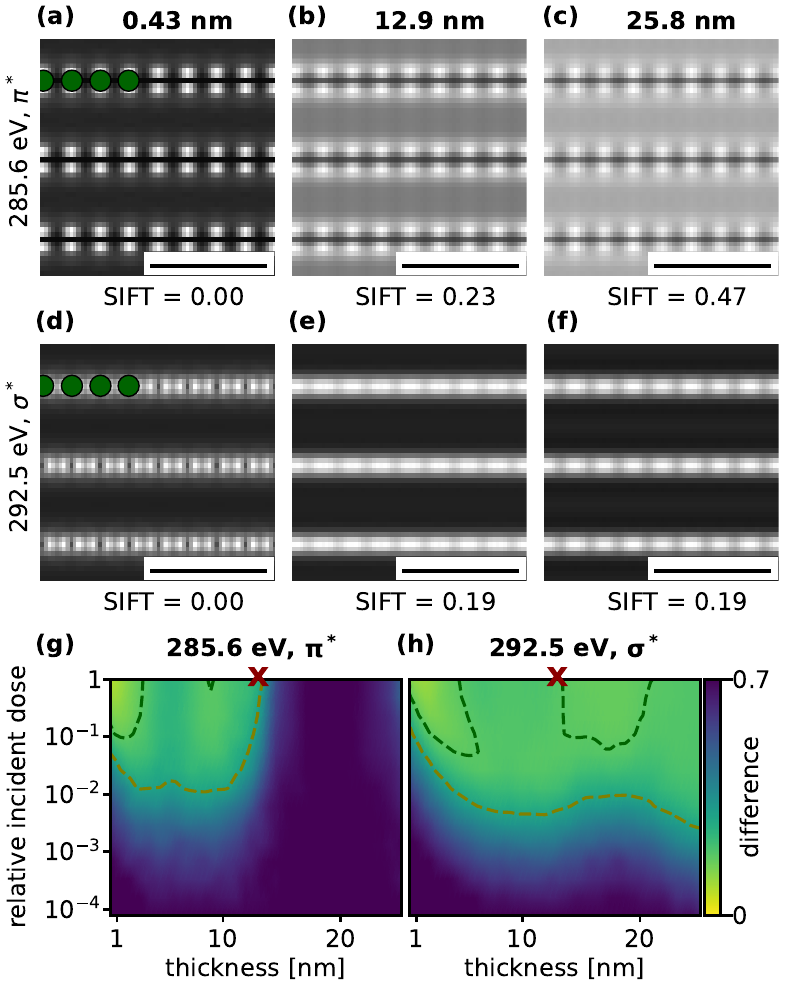}
    \caption{Exactly the same as Fig.~\ref{fig_STEM_graphite_s2n} but with simulated EFTEM images.
}
    \label{fig_TEM_graphite_s2n}
\end{figure}

\begin{figure}
    \centering
    \includegraphics{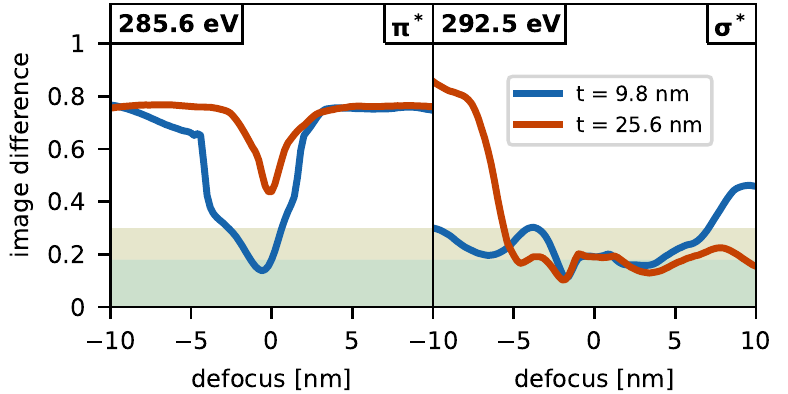}
    \caption{SIFT image difference of EFTEM images as a function of defocus for graphite sample thicknesses of 9.8~nm and 25.6~nm.
    All image differences are relative to the orbital map for 0 defocus.
}
    \label{fig_TEM_graphite_defocus}
\end{figure}

In the following, we present the EFTEM orbital maps of the graphite $\pi^\ast$ and $\sigma^\ast$ orbitals simulated with parameters corresponding to those of the STEM spectrum images in the previous subsection.
While the basic shape of the orbitals in the reference images for a one unit cell thick sample is in principle the same as in the STEM images, there are some distinctions that influence the SIFT image differences.
The orbitals in EFTEM for both energy losses appear more sharply and with clearer edges.
Further, for the $\pi^\ast$ orbitals, there is no intensity at all directly at the positions of the carbon atoms, the two lobes of the p orbital are distinctly separated. 
This was also the case for all investigated sample thicknesses. 
Further, an increase in sample thickness results in a large intensity increase between the atomic columns. 
This effect continues up to the point where the background intensity is nearly as high as the intensity of the orbitals themselves.
For finite incident electron dose, this effect is especially pronounced with the result of a very large image difference. 
The image difference map in Fig.~\ref{fig_TEM_graphite_s2n}(g) shows a steep jump in image difference around 14~nm, from which thickness on the features of the orbitals are no longer discernible from the background.\\
For the $\sigma^\ast$ orbital the preferred inelastic scattering direction is mainly in the plane of the carbon sheets, leading to the opposite result compared to the $\pi^\ast$ case. 
The circular shapes at the positions of the carbon atoms are fully smeared out in the horizontal direction for samples exceeding 5~nm. 
We want to emphasize here that this effect is a fundamental result of the electron propagation through the material, i.e., it cannot be overcome with an increase in spatial or spectral resolution, only with a thinner sample.
The thickness -- dose difference map, however, reveals an oscillatory thickness dependence, similar to the rutile EFTEM images. 
Due to this oscillation, a sample thickness between 18~nm and 22~nm again nearly allows to distinguish individual orbitals.
\noindent 
As for rutile, we have investigated the dependence of the image difference on the defocus (Fig.~\ref{fig_TEM_graphite_defocus}).
For the energy loss of 285.6~eV, no image improvement is achieved by either under- or overfocusing. 
On the contrary, for relatively thick samples of 25.6~nm, already a small defocus of 1~nm in either direction is detrimental for the orbital image.
This again highlights the tremendous precision required for orbital mapping.
For the energy loss of 292.5~eV, we see, similar to the rutile EFTEM images, a significant improvement of image quality for a small (between $-5$ and $-2$ nm) defocus.
In this case, however, the sample thickness has little influence on the effect of the defocus on the image.

%% file: STOLMO.tex
\subsection{STO-LMO}
The last structure of our parameter investigation is a heterostructure consisting of the two perovskites \ce{SrTiO3}~(STO) and \ce{LaMnO3}~(LMO).
Perovskites are a structure model of ever more increasing public interest~\cite{Bhalla2000, Fiebig2005} due to their extreme range in synthesizable properties like superconductivity, ferroelectricity and ferromagnetism~\cite{Liu2017}.
The combination of thin layers of different perovskites to a heterostructure allows the already widespread application of perovskites as lasers, LEDs, photodetectors and solar cells~\cite{Cheng2022, Ansari2018, Snaith2013, Yin2014,Kazim2014, Niu2015}. \\
The cubic perovskite STO has the spacegroup~\hmn{Pm-3m} with lattice parameter $a_{\mathrm{STO}} = 3.905$~\AA{}~\cite{Mitchell2000}. 
In contrast, LMO displays rhombohedral symmetry according to the \hmn{R-3c} space group with lattice parameters $a_{\mathrm{LMO}} = 5.508$~\AA{}  and $c_{\mathrm{LMO}} = 13.310$~\AA{}~\cite{Dezanneau2003}. 
While the structure is very similar to the cubic perovskite structure, the \ce{MnO6} octahedra are rotated a few degrees relative to each other, reducing the crystal symmetry~\cite{Glazer1972}.
For epitaxial deposition on a STO substrate, however, LMO is formed in a pseudo-cubic perovskite structure with an effective lattice parameter $a_{\mathrm{effLMO}} = 3.914$~\AA{}~\cite{Verbeeck2002, Norby1995}.
Due to the almost perfect match between the structure parameters (0.2~$\%$ mismatch), a stable heterostructure can be constructed.
Computational constraints limit us to a periodic structure of 4~unit cells of each of the materials.
Both, a \ce{Ti}--\ce{La} terminated interface and a \ce{Mn}--\ce{Sr} terminated interface are present in the structure (presented in supplementary information).
However, we will exclusively focus on the \ce{Ti}--\ce{La} interface including the first adjacent unit cells, thus the supercell size is more than sufficient.\\
Due to the presence of oxygen in both perovskites, we have selected the \ce{O} K-edge for orbital mapping.
In particular, two energy losses are of interest for the \hkl[1 0 0] zone axis.
At 533.4 eV, the imaged orbitals of the specific \ce{O} atoms located between \ce{Sr} atoms display an orientation perpendicular to the interface, which we will define as p$_x$ orbitals in this work. 
The corresponding LMO \ce{O} atoms located between \ce{La} atoms, however, display an orthogonal direction, which we will define as p$_y$ orbitals.
These specific orbitals are of special interest as they, contrary to most of the remaining \ce{O} orbitals, display a consistent orientation over all unit cells of the corresponding material and over energy windows as large as 1~eV (Fig.~S4 in the SI).
At the second chosen energy loss, however, at 536.7~eV, the orbitals swap their orientation with each other. 
The previous p$_x$ orbitals of STO have changed to p$_y$ orbitals and vice versa for LMO.
Even though the mentioned orbitals are of particular interest to us, we will present EFTEM and STEM spectrum images of all the orbitals in close proximity to the \ce{Ti}-\ce{La} terminated interface.
Likewise, the SIFT image difference will take into account all the orbitals.

%Perovskites
%multiferroics with strong magnetoelectric coupling at room temperature are the most promising materials for data storage cite(Liu)
%planar heterojunction important, relying on the ambipolar characteristics of perovskite thin films cite(LiuNature)
%corner-shared oxygen octahedra linked in the three dimensions is the basic feature of the perovskite feature cite(Bhalla)
%ferroelectricity result of the Octahedra
%family of perovskites called "relaxors" with extraordinary high dielectric constants (Bhalla) 
%microwave dielectrics, high T$_C$ superconductors 

\subsubsection{STEM}
\begin{figure}
    \centering
    \includegraphics{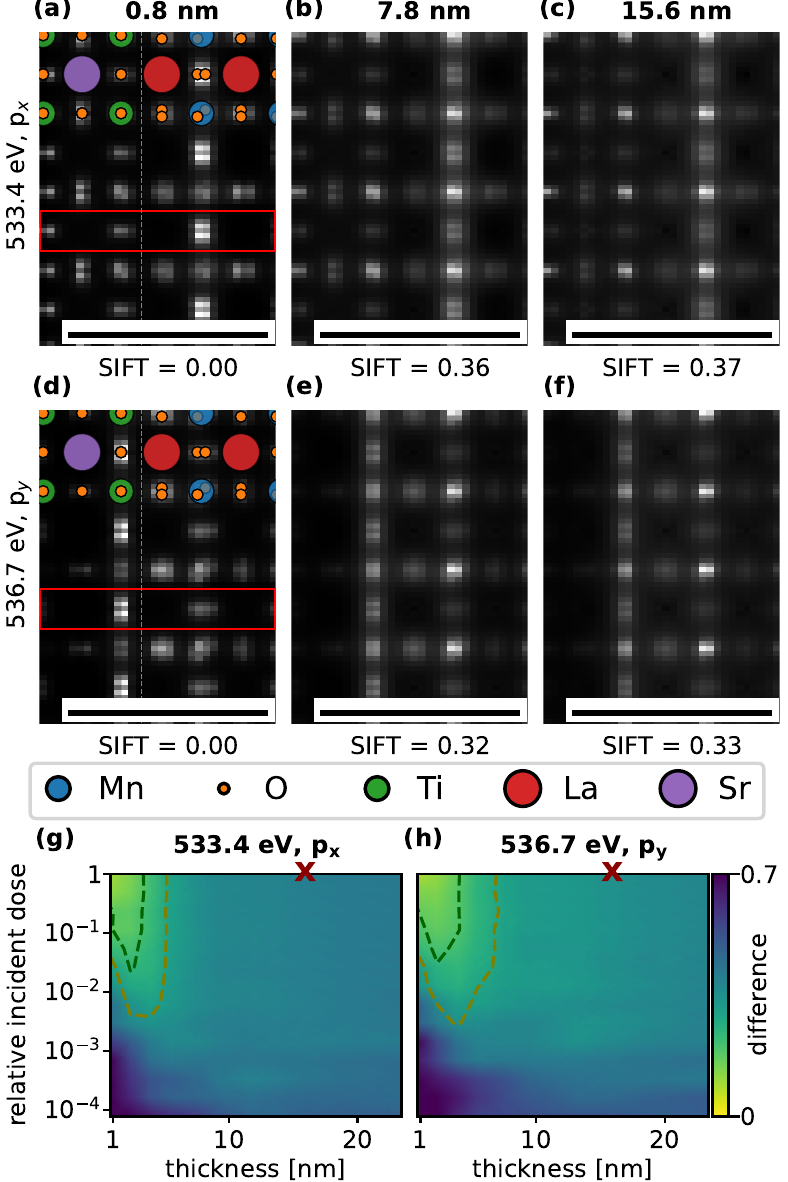}
    \caption{(a) -- (f) Simulated STEM spectrum images of the interface between STO and LMO in the \hkl[1 0 0] zone axis for both materials, the indicated sample thickness and energy loss with an infinite electron dose.
    All scale bars indicate 1~nm. 
    (g), (h) Image difference maps of the STO-LMO spectrum images for the indicated energy losses.
    The incident dose is measured relative to the image dose of $10^6$~e$^-$/nm$^2$ for the image of a 15.6~nm thick sample, marked by the red x.
    The reference image in all cases is calculated with infinite electron dose.
    The dark green and olive contour lines indicate an image difference of 0.18 and 0.3, respectively.
}
    \label{fig_STEM_STOLMO_s2n}
\end{figure}

\begin{figure}
    \centering
    \includegraphics{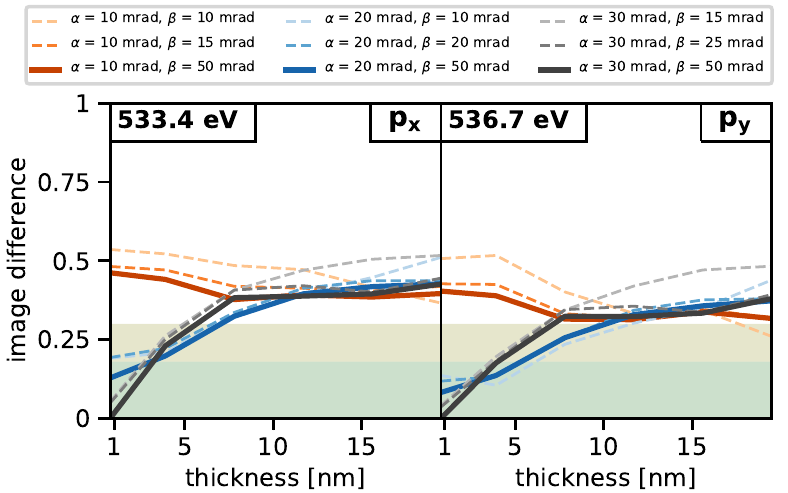}
    \caption{STO-LMO SIFT image difference as a function of sample thickness for an acceleration voltage of 300~kV and several convergence and collection semi-angles $\alpha$ and $\beta$. 
    All image differences are relative to the orbital map for $\alpha$ = 30~mrad and $\beta$ = 50~mrad. 
}
    \label{fig_STEM_STOLMO_conv}
\end{figure}

The resulting STEM spectrum images are shown in Fig.~\ref{fig_STEM_STOLMO_s2n}. 
The upper row in Fig.~\ref{fig_STEM_STOLMO_s2n} displays the result for an energy loss of 533.4~eV.
There is a clear distinction between the two parts of the heterostructure.
In the region of interest, marked by the red rectangle, the p-like orbitals are oriented perpendicular to each other. 
Another noticeable difference is that in the LMO region, all observed orbitals are blurry and less distinct. 
This is a direct consequence of the rotation of the \ce{MnO6} octahedra.
The slightly varying projected positions of the \ce{O} atoms lead to an overlay of orbital images and, thus, less distinct shapes.
Increasing the sample thickness results in a relative intensity increase at the positions of the heavier \ce{Ti} and \ce{Mn} atoms 
due to channeling effects.
Further, at these positions, the originally p-like shapes deteriorate to a circularly symmetric shape.
Thus, all potential extractable directional orbital information is lost.
On the other hand, this relative intensity increase leads to an intensity decrease in our region of interest and the STO p$_x$ orbitals have nearly vanished for a sample thickness larger than $~10$~nm.\\
The situation is very similar for the energy loss of 536.7~eV.
While the effects of channeling along \ce{Ti} and \ce{Mn} columns are just as present, there are a few differences.
As mentioned before, most of the orbitals have switched their orientation, in particular all of the orbitals in the marked region.
However, of the STO p$_y$ orbitals, only the one closest to the interface is visible.
Nevertheless, for both energy losses, the majority of details and orbital information vanishes for increasing sample thickness.
The thickness--dose plots [Fig.~\ref{fig_STEM_STOLMO_s2n}(g) and (h)] further confirm this. 
Although the image difference appears to reach a plateau for all incident doses above 8--10~nm, the difference values are in both cases above the acceptable threshold and, in general, significantly higher than for the previous materials.
Increasing or decreasing the electron beam convergence semi-angle $\alpha$ appears to have no significant influence on the orbital map quality for a sample thickness above 10~nm (Fig.~\ref{fig_STEM_STOLMO_conv}).
This is in line with the $\alpha$ dependence of STEM orbital maps for rutile (Fig.~\ref{fig_STEM_rutile_conv}) and the $\sigma^\ast$ orbital of graphite [Fig.~\ref{fig_STEM_graphite_conv}(b)], although for STO-LMO, the effect is even more pronounced.

\noindent

\subsubsection{TEM}
\begin{figure}
    \centering
    \includegraphics{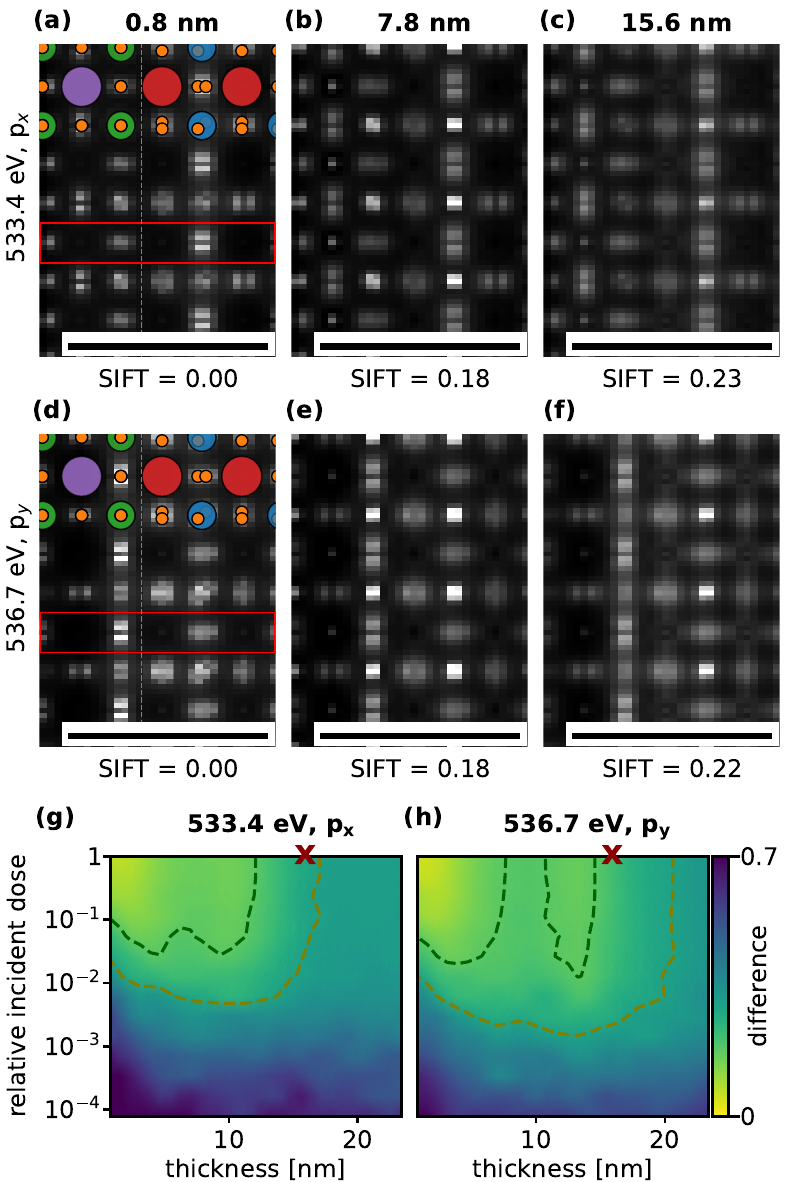}
    \caption{The same as in Fig.~\ref{fig_STEM_STOLMO_s2n} but with simulated EFTEM images.
}
    \label{fig_TEM_STOLMO_s2n}
\end{figure}

\begin{figure}
    \centering
    \includegraphics{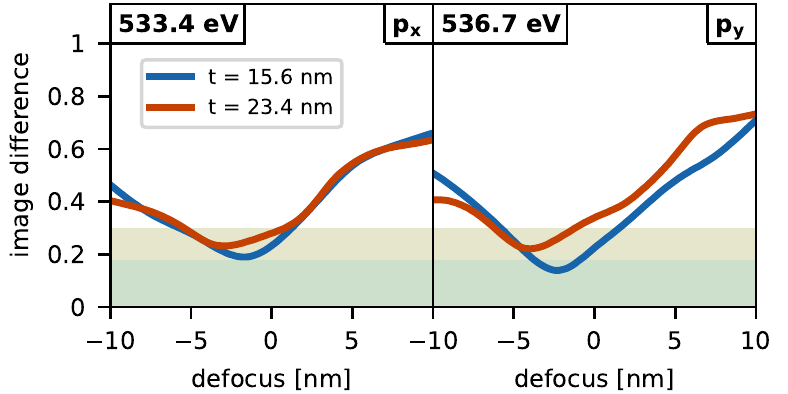}
    \caption{SIFT image difference of EFTEM images as a function of defocus for an STO-LMO interface with sample thicknesses of 15.6~nm and 23.4~nm.
    All image differences are relative to the orbital map for 0~defocus.
}
    \label{fig_TEM_STOLMO_defocus}
\end{figure}

The EFTEM images for the STO-LMO heterostructure are presented in Fig.~\ref{fig_TEM_STOLMO_s2n} and are remarkably similar to the STEM spectrum images presented in the previous section.
However, while an increasing sample thickness still shows a relative intensity shift to the \ce{Ti} and \ce{Mn} sites, it is less severe than in the STEM case.
The resulting image difference numbers are relatively low, even for realistically thick samples.
The thickness--dose plots [Fig.~\ref{fig_TEM_STOLMO_s2n}(g) and (h)] show acceptable differences up to 17~nm for 533.4~eV and up to 20~nm for 536.7~eV, as long as a high enough electron dose is used. 
In the latter case, the effect of the Pendellösung can be clearly observed, in the same way as for the rutile and graphite EFTEM images before.
There is a periodicity in the image difference dependent on the sample thickness which leads to an "island" of low image difference from 11~nm to 13~nm. 
Taking these effects into account, it seems that for this particular structure, EFTEM is more suited to the task from a purely image forming point of view.
A small defocus of $-1$~nm to $-3$~nm can further improve the image quality, following the same apparent pattern as the other investigated materials but with less ambivalence in this case (Fig.~\ref{fig_TEM_STOLMO_defocus}).

\noindent

